\documentclass[a4paper,12pt]{article}
\pdfoutput=1
\usepackage{geometry}
\usepackage{amsmath,amssymb,amsfonts}
\usepackage{xcolor,graphicx,cite,soul}
\usepackage{array,booktabs,longtable}
\usepackage{caption,microtype}
\usepackage{subcaption}
\usepackage{hyperref}
\usepackage{bm}
\usepackage{bbm}
\hypersetup{
  colorlinks   = true, 
  urlcolor     = blue, 
  linkcolor    = black, 
  citecolor   = blue 
}
\usepackage[numbers,sort&compress]{natbib}
\let\OLDthebibliography\thebibliography
\renewcommand\thebibliography[1]{
  \OLDthebibliography{#1}
  \setlength{\parskip}{4pt}
  \setlength{\itemsep}{0pt plus 0.3ex}
}

\allowdisplaybreaks
\usepackage{multicol}
\usepackage{enumitem}

\usepackage{datetime}
\usepackage{appendix}
\usepackage[normalem]{ulem} 
\usepackage{multirow}
\usepackage{cancel}

\makeatletter
\providecommand{\tabularnewline}{\\}

\makeatother

\newcommand{\lsim}
{\;\raisebox{-.3em}{$\stackrel{\displaystyle <}{\sim}$}\;}

\newcommand\al{\alpha}
\newcommand\be{\beta}
\newcommand\tb{\tan\beta}
\newcommand\TB{t_\beta}

\newcommand\CTb{\cot\be}

\newcommand\CBA{c_{\beta - \alpha}}
\newcommand\SBA{s_{\beta - \alpha}}

\newcommand\ReDiag{\mathop{%
  \raise .5pt\hbox{[}%
  \widetilde{\mathrm{Re}}%
  \raise .5pt\hbox{]}}}
\newcommand\ReOffDiag{\mathop{%
  \raise .5pt\hbox{$\llbracket$}%
  \widetilde{\mathrm{Re}}%
  \raise .5pt\hbox{$\rrbracket$}}}

\newcommand\Mh{m_h}
\newcommand\MH{m_H}
\newcommand\MA{m_A}
\newcommand\MHp{m_{H^\pm}}

\newcommand\msq{m_{12}^{2}}

\newcommand\refeq[1]{Eq.~(\ref{#1})}

\newcommand\refta[1]{Tab.~\ref{#1}}
\newcommand\refse[1]{Sect.~\ref{#1}}

\newcommand\citere[1]{Ref.~\cite{#1}}
\newcommand\citeres[1]{Refs.~\cite{#1}}

\newcommand{\CP}{{\cal CP}}
\newcommand{\cp}{{\CP}}

\newcommand{\gev}{\,\, \mathrm{GeV}}

\newcommand\fb{\ensuremath{\,\mbox{fb}}}
\newcommand\ab{\ensuremath{\,\mbox{ab}}}

\newcommand\ifb{\ensuremath{\fb^{-1}}}
\newcommand\iab{\ensuremath{\ab^{-1}}}

\newcommand\mh[1]{m_{h_{#1}}}

\newcommand{\br}{\text{BR}}

\def\reffi#1{\mbox{Fig.~\ref{#1}}}
\def\reffis#1{\mbox{Figs.~\ref{#1}}}

\def\la{\lambda}

\newcommand{\lahhh}{\ensuremath{\la_{hhh}}}
\newcommand{\lahhH}{\ensuremath{\la_{hhH}}}

\newcommand{\xila}{\ensuremath{\xi_H^t \times \lahhH}}
\newcommand{\xilaNN}{\xila {\rm (NN\;predicted)}}
\newcommand{\xilaTH}{\xila {\rm (theory)}}

\newcommand{\mhh}{\ensuremath{m_{hh}}}

\definecolor{Orange}{named}{orange}
\definecolor{Purple}{named}{purple}
\definecolor{Lightblue}{cmyk}{0.9,0.1,0.1,0.3}
\definecolor{dgelborange}{cmyk}{0.,0.3,0.5, 0.}
\definecolor{Lila}{rgb}{0.5,0.,1}
\definecolor{Darkgreen}{rgb}{0.,.7,0.2}

\newcommand{\htr}[1]{{\color{red} #1}}

\graphicspath{{figs/}}
\allowdisplaybreaks
\begin{document}
\thispagestyle{empty}

\def\thefootnote{\fnsymbol{footnote}}

\begin{flushright}
\mbox{}
DESY-25-087\\
IFT--UAM/CSIC-25-067 \\
KA-TP-19-2025 
\end{flushright}


\begin{center}

{\large\sc 
{\bf 
Experimental Determination of BSM Triple Higgs Couplings\\[.5em] 
at the HL-LHC with Neural Networks
 }}  

\vspace{1cm}

{\sc
Markus Frank$^{1}$%
\footnote{markus.frank@x-ent.tech}%
, Sven Heinemeyer$^{2}$%
\footnote{Sven.Heinemeyer@cern.ch}%
, Margarete M\"uhlleitner$^{3}$%
\footnote{margarete.muehlleitner@kit.edu}\\[.5em]
~and Kateryna Radchenko$^{4}$%
\footnote{kateryna.radchenko@desy.de}%
}
\vspace*{.7cm}

{\sl
$^1$Independent Researcher, Karlsruhe, Germany

\vspace*{0.1cm}

$^2$Instituto de F\'isica Te\'orica (UAM/CSIC), 
Universidad Aut\'onoma de Madrid, \\ 
Cantoblanco, 28049, Madrid, Spain

\vspace*{0.1cm}

$^3$Institute for Theoretical Physics,
Karlsruhe Institute of Technology, 76128 Karlsruhe, Germany

\vspace{0.1em}

$^4$Deutsches Elektronen-Synchrotron DESY, Notkestr.\ 85, 22607 Hamburg,
Germany

}

\end{center}

\vspace*{0.1cm}

\begin{abstract}
\noindent
The shape of the Higgs potential is modified by the presence of additional scalar fields, as predicted in many Beyond-Standard-Model (BSM)
scenarios. In such cases, deviations in the Higgs self-interactions — in particular the trilinear Higgs couplings — could serve 
to disentangle the physics beyond the Standard Model (SM). While the SM predicts only one trilinear Higgs coupling, extended scalar sectors allow for additional 
self-interactions that can manifest themselves in Higgs pair production, via the $s$-channel contribution of a heavy $\cp$-even scalar $H$. We present the first sensitivity study to such a BSM trilinear scalar coupling using machine learning. Specifically, we train a neural 
network on the invariant mass distributions of Higgs pair production at the HL-LHC to extract \xila, i.e.~the product of the resonant~$H$ top-Yukawa
coupling and the trilinear coupling of $H$ to the two SM-like Higgses in the final state, $hh$. Assuming a hypothetical $H$ mass of 450~GeV, we show 
that, depending on future experimental efficiencies and uncertainties, a determination of \xila\ at the 10–20\% level may be achievable by the end
of the HL-LHC. We present a simple and more efficient alternative to classical statistical methods, proving the efficiency of neural networks for 
both hypothesis testing and parameter estimation, which outperforms conventional maximum likelihood methods in this context.

\end{abstract}


\def\thefootnote{\arabic{footnote}}
\setcounter{page}{0}
\setcounter{footnote}{0}

\newpage


\section{Introduction}
\label{sec:intro}
The discovery of a Higgs boson, $h$, at ATLAS and CMS in 2012, nearly 50 years
after its prediction, was a milestone in high-energy physics~\cite{Aad:2012tfa,Chatrchyan:2012xdj}.
Within theoretical and experimental uncertainties this new particle is
consistent with the existence of a Standard-Model~(SM) Higgs boson at a mass
of~$\sim 125 \gev$~\cite{Khachatryan:2016vau}. Now
the reconstruction of the Higgs potential is one of the main goals in particle physics in the years ahead.
A primary objective is to establish the trilinear Higgs coupling (THC), $\lahhh$, with a high level of precision. 
This task can partially be accomplished in the High Luminosity phase of the Large Hadron Collider 
(HL-LHC)~\cite{deBlas:2019rxi}. 
While no sign of Beyond-Standard-Model (BSM) physics was yet discovered at the LHC, the measurements of Higgs-boson production and decay rates, which are known experimentally to a precision of roughly $\sim 10-20\%$ \cite{ATLAS:2022vkf, CMS:2022dwd},  
leave ample room for BSM interpretations \cite{Abouabid:2021yvw}. Consequently,
the possibility of additional scalar states present in nature that have escaped detection so far remains an enticing research avenue. If a new signal is established in the current or upcoming runs of the LHC, the next question will
be to determine the couplings of the new scalar states at the highest possible level of precision. In particular, the THCs 
of these new scalar states are crucial for a full reconstruction of the Higgs potential.

One important reason to consider a non minimal scalar sector is the possibility to accommodate a strong first order
electroweak phase transition (SFOEWPT), a necessary condition for electroweak baryogenesis ~\cite{Cline:1996mga,Fromme:2006cm,
Cline:2011mm,Dorsch:2016nrg}. 
It has been found that the SM does not feature such a transition with the mass of the discovered Higgs being 
$\sim 125 \gev$~\cite{Kajantie:1996mn}. However, models with extended scalar sectors 
can accommodate a SFOEWPT, cf.~e.g.~for models with a second Higgs doublet~\cite{Cline:1996mga, Fromme:2006cm,Dorsch:2013wja,Dorsch:2014qja, Dorsch:2016nrg, 
Basler:2016obg,Dorsch:2017nza,Basler:2017uxn,Goncalves:2021egx,Biekotter:2022kgf}, which in
turn may lead to a detectable gravitational waves signal \cite{Caprini:2015zlo, Caprini:2019egz,Caprini:2024hue}. 

A very well motivated extension of the SM that could yield such a signal is the 
Two-Higgs-Doublet Model (2HDM)~\cite{TDLee,Gunion:1989we,Aoki:2009ha,Branco:2011iw},
that extends the field content of the SM by an additional complex Higgs doublet.
This results after electroweak symmetry breaking in five physical scalar fields, two $\cp$-even scalar fields,
$h$ and $H$, where by convention $\Mh < \MH$ (we assume $\mh\ \sim 125 \gev$), 
one $\cp$-odd field, $A$, and one charged Higgs pair, $H^\pm$.
If such a scenario was realized in nature, not only the measurement of the SM-like THC, $\lahhh$, would be needed,
but also experimental sensitivity to the remaining THCs among all the scalars in the theory 
will be necessary to determine the shape of the Higgs potential.


Triple Higgs couplings enter at leading order in the main process of di-Higgs production at the LHC, given by gluon fusion. 
Consequently, this process is of particular interest in the experimental determination of the THCs.
Within the SM, two Feynman diagrams contribute to $hh$ production: via a top-quark triangle a light Higgs, $h$, is produced
in the $s$-channel with the subsequent decay into $hh$, 
and the diagram where the two Higgs bosons are radiated from a top-quark box.
In the 2HDM there is one additional contribution to this process: via a top-quark triangle a heavy Higgs,
$H$, is resonantly produced. Consequently, $hh$ production would provide access not only to $\lahhh$, 
but also to $\lahhH$. The relevant quantity for the $H$ contribution is $\xi_H^t \times \lahhH$, where
$\xi_H^t$ parametrizes the top-Yukawa coupling of the heavy Higgs boson.

In view of an ever increasing availability of experimental data, ``old school'' data treatment such as $\chi^2$ tests 
become insufficient when dealing with the determination of BSM THCs at HL-LHC.
The broad selection of newly available tools for data analysis has been used extensively by the high-energy
physics community in the past years \cite{Plehn:2022ftl, Feickert:2021ajf}. 
In particular, in experimental particle physics, modern machine learning techniques have proven to be extremely
useful in, e.g., the event selection. 
In particle physics phenomenology a wide range of studies has also started to employ neural networks 
and machine learning techniques as a tool to learn about the underlying physics of nature.

In this work, we will study the possibility of inferring the magnitude of \xila\ from the 
invariant mass distribution of the di-Higgs pair, $\mhh$, in the process $gg \to hh$. To this end, we will analyze the performance of machine
learning techniques, more specifically a neural network (NN), in the determination of such couplings using 
anticipated HL-LHC data for di-Higgs production, as opposed to the conventional statistical data analysis methods 
that are insufficient for the extraction of BSM THCs~\cite{Arco:2022lai}. 
To the best of our knowledge, this work constitutes the first analysis attempting to
quantitatively extract BSM THCs from (HL-)LHC data.

This paper is organized as follows. First, in Sec.~\ref{sec:model}, we give a very brief introduction of the 2HDM where we settle our notation and specify the analyzed parameter region as well as the constraints applied in the model. In Sec.~\ref{sec:stat} we provide a classical statistical analysis of the data to determine \xila. In Sec.~\ref{sec:method} we give details of the methodology followed in our NN  analysis, in particular in the chosen data sets and type of network. We show our results in Sec.~\ref{sec:results}, where a comparison between both methods and future prospects for the determination of $\xila$ are analyzed. We state our conclusions in Sec.~\ref{sec:conclusions}. 

\section{The 2HDM}
\label{sec:model}

Our analysis is performed in the $\cp$-conserving 2HDM with a softly broken $\mathbb{Z}_2$
symmetry~\cite{TDLee,Gunion:1989we,Aoki:2009ha,Branco:2011iw}. This discrete symmetry is imposed in order to avoid 
flavor changing neutral currents at tree level, which are experimentally found to be very suppressed.
The 2HDM Higgs sector is comprised of two complex Higgs doublets, leading after EWSB to five physical Higgs bosons, 
two $\cp$-even states $h$ and $H$, one $\cp$-odd state $A$ and two charged states $H^{\pm}$. 
The mass hierarchy of the $\cp$-even particles is by convention $\MH >\Mh$, and we choose to identify
the light state $h$ with the experimentally measured state at $\approx 125 \gev$. 
The electroweak precision observables constrain 
the masses of the remaining scalars such that either $\MHp \approx \MH$ or 
$\MA \approx \MH$. Therefore, for simplicity, in the following analysis we will assume 
$m_{\phi} \equiv \MH=\MA=\MHp>\Mh$. (As will become clear later, the masses of the $\cp$-odd and charged Higgs bosons
play a negligible role in our analysis.) The imposed $\mathbb{Z}_2$ symmetry is allowed to be softly broken by a 
dimensionful mass term $m_{12}^2$, which is a free parameter of the model. 
Furthermore, there are two mixing angles $\alpha$ and $\beta$ that diagonalize the $\cp$-even and 
$\cp$-odd/charged sectors, respectively. It is convenient to use the parameterization $\tb \equiv\TB$, which is the
ratio of the vacuum expectation values of the two doublets, i.e.~$\TB = v_2/v_1$, and $\CBA \equiv \cos(\beta-\alpha)$, 
which is zero in the so-called alignment limit \cite{Gunion:2002zf}.
In this limit all the couplings of the scalar $h$ to fermions 
and gauge bosons have the same values as predicted in the SM and in particular $\lahhH = 0$.

In this set-up, the free parameters describing the tree-level Higgs sector of the model, given that $\Mh\approx 125 \gev$ and $v=\sqrt{v_1^2+v_2^2} \approx 246 \gev$, can be chosen as

\begin{equation}
    m_{\phi},\; m_{12}^2,\; \TB,\; \CBA~.
\end{equation}

The imposition of a discrete $\mathbb{Z}_2$ symmetry leads four different 2HDM types,
depending on the extension of the $\mathbb{Z}_2$ symmetry to the fermion sector.  This determines the Yukawa couplings 
of the Higgs bosons to the fermions. In this work we will focus only on scenarios in the 
so called ``Type~I" 2HDM, although the type of the 2HDM plays a subleading role and we can safely generalize our results, e.g.~the parameter in which we 
are interested, 
$\xi_H^t \times\lambda_{hhH}$, is the same in all four Yukawa types.  
Relevant to our analysis here will be the Yukawa couplings of the $\cp$-even Higgs bosons, 
which enter in the 2HDM Lagrangian as
\begin{equation}
	\mathcal{L} =-\sum_{f=u,d,l}\frac{m_f}{v}\left[\xi_h^f\bar{f}fh + \xi_H^f\bar{f}fH \right],
\end{equation}
where $m_f$ are the fermion masses and $\xi_{h,H}^f$ are the fermionic Yukawa coupling modifiers, 
which in Type~I are equal for all three generations of up-type quarks, down-type quarks and leptons. Since the Higgs couples most strongly to the top quark, we will consider only the top Yukawas in the following
$(f = t)$. In all four types the Yukawa coupling modifiers for the top quark are given by 
\begin{equation}
\xi_{h}^{t} = \SBA + \CBA \CTb, \;\;\; \xi_{H}^{t} = \CBA - \SBA \CTb.
\end{equation}
And the tree level expressions of the THCs (which are independent of the Yukawa type) are given by
\begin{align}
  \lahhh
        &= \frac{1}{2v^2} \bigg\{ \Mh^2 s_{\be -\al }^3 
        +\left(3 \Mh^2-2 M^2\right) c_{\be -\al }^2 s_{\be -\al }
        +2 \cot 2 \be \left( \Mh^2-M^2\right) c_{\be -\al
        }^3\bigg\}\;, \\[.5em] 
\notag
 \lahhH
        &= -\frac{\CBA}{2v^2} \bigg\{  \left(2\Mh^2+\MH^2-4M^2\right)\SBA^2
        +2\cot{2\be}\left(2\Mh^2+\MH^2-3M^2\right) \SBA\CBA \\
        &\quad\qquad -\left(2\Mh^2+\MH^2-2M^2 \right) \CBA^2
         \bigg\}\;,  
\notag
\end{align}
\noindent 
where $\SBA \equiv\sin(\beta-\alpha)$ etc. and $M^2$ is defined as
\begin{eqnarray} 
 M^2&=&\frac{\msq}{s_\be c_\be} \,.
\label{eq:mbar}
\end{eqnarray} 
It should be noted here that we do not consider loop corrections to the THC, although these 
can be relevant in di-Higgs production~\cite{Kanemura:2002vm,Kanemura:2004mg,Nhung:2013lpa,Borschensky:2022pfc,Heinemeyer:2024hxa}. We focus on the question whether,
given the predictions for the invariant mass distribution in Higgs pair production through gluon fusion within a theoretical framework, the experimental determination of the couplings is possible.
In the hypothetical case of a future detection of a heavy Higgs resonance in di-Higgs production,
such a framework must be improved with higher order corrections. 
However, this is not expected to alter the performance of the NN analysis.


\subsection{Higgs Pair Production}
\label{sec:hh-2hdm}

\begin{figure}[ht!]
  \begin{center}
	   \includegraphics[width=1\linewidth]{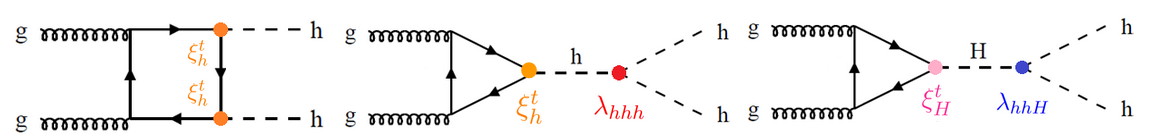}
  \end{center}
\caption{Leading-order diagrams to 2HDM SM-like Higgs pair production in the gluon fusion process at hadron colliders.}  
\label{fig:diagrams}
\end{figure}

In the framework of the 2HDM,  Higgs pair production has been widely studied in the literature for many years
(cf.~e.g.~\citere{Plehn:1996wb,Dawson:1998py,Djouadi:1999rca} for the Minimal Supersymmetric Standard Model (MSSM) that shares the 
2HDM Higgs sector with Yukawa Type~II, and e.g.~\citeres{ Arhrib:2009hc,Asakawa:2010xj,Grober:2017gut,Basler:2017uxn,Basler:2019nas,Abouabid:2021yvw,  Arco:2022lai, Baglio:2023euv,Heinemeyer:2024hxa} for the 2HDM). 
%
The leading-order diagrams for gluon fusion into a SM-like Higgs pair in the 2HDM are shown in \reffi{fig:diagrams}. The box (left) and the
$h$ exchange in the $s$-channel (middle) comprise the continuum contribution that is present in the SM 
(modulo changes in the Higgs-boson couplings w.r.t.\ their SM values). The resonant diagram (right) involves 
the heavy $\cp$-even Higgs-boson, $H$, exchanged in the $s$-channel. The cross section given by the continuum
diagrams is proportional to the SM-like top Yukawa coupling $\xi_h^t$ and the trilinear Higgs self-interaction
$\lahhh$. The resonant diagram, entering in the full calculation of the di-Higgs production cross
section in the 2HDM is proportional to the product of the heavy Higgs-boson top-Yukawa coupling
$\xi_H^t$ and the THC $\lahhH$. The interference of this diagram with the continuum contribution 
creates a dip-peak or peak-dip structure in the $\mhh$ distribution, located at $\sim \MH$, where the type of the
structure depends on the sign of the product of (\xila)~\cite{Arco:2022lai}.
It was furthermore shown in \citere{Arco:2022lai} that in parameter regions close to the alignment limit, i.e.\ with 
the properties of the light $\cp$-even Higgs boson in good agreement with the LHC rate measurements, and with 
$\MH \lsim 1000 \gev$ the resonant $H$ contribution can modify the $\mhh$ distribution substantially. This suggests
that an extraction of \xila\ via the measurement of the invariant mass distribution \mhh\ of the di-Higgs production cross section may be possible.

For the theoretical prediction of the Higgs pair production cross section we use
the code {\tt HPAIR} \cite{Plehn:1996wb,Dawson:1998py} adapted for the 2HDM,
which also computes the invariant mass distribution of the two light Higgses $hh$ in the final state, \mhh. It allows to include next-to-leading-order QCD corrections in the heavy-top limit.


\subsection{Model Constraints}
\label{sec:constraints}

In order to analyze viable 2HDM parameter points, the relevant constraints on the model have to be
taken into account. In this work we have applied the most relevant constraints coming from theoretical considerations and previous
experimental measurements.
In order to take these constraints into account we have used the public tool 
\texttt{thdmTools}~\cite{Biekotter:2023eil}. In particular, we have applied bounds for perturbative unitarity 
(we choose an upper bound for the eigenvalues of the scalar four-point scattering matrix of $8\pi$), 
vacuum stability, constraints from electroweak precision observables, flavor constraints and 
constraints from direct BSM Higgs-boson searches at the LHC, Tevatron and LEP, as well as the
compatibility with the signal strength of the Higgs-boson at $\sim 125 \gev$. The latter two are taken
into account through the interface to \texttt{HiggsTools}~\cite{Bahl:2022igd} (which merged and updated \texttt{HiggsBounds}~\cite{Bechtle:2008jh, Bechtle:2011sb,Bechtle:2013wla,Bechtle:2015pma,Bechtle:2020pkv} and \texttt{HiggsSignals}~\cite{Bechtle:2013xfa, Bechtle:2014ewa,Bechtle:2020uwn}).


\subsection{Benchmark scenario}
\label{sec:benchmark}

\begin{figure}[ht!]
\centering
\includegraphics[width=0.54\textwidth]{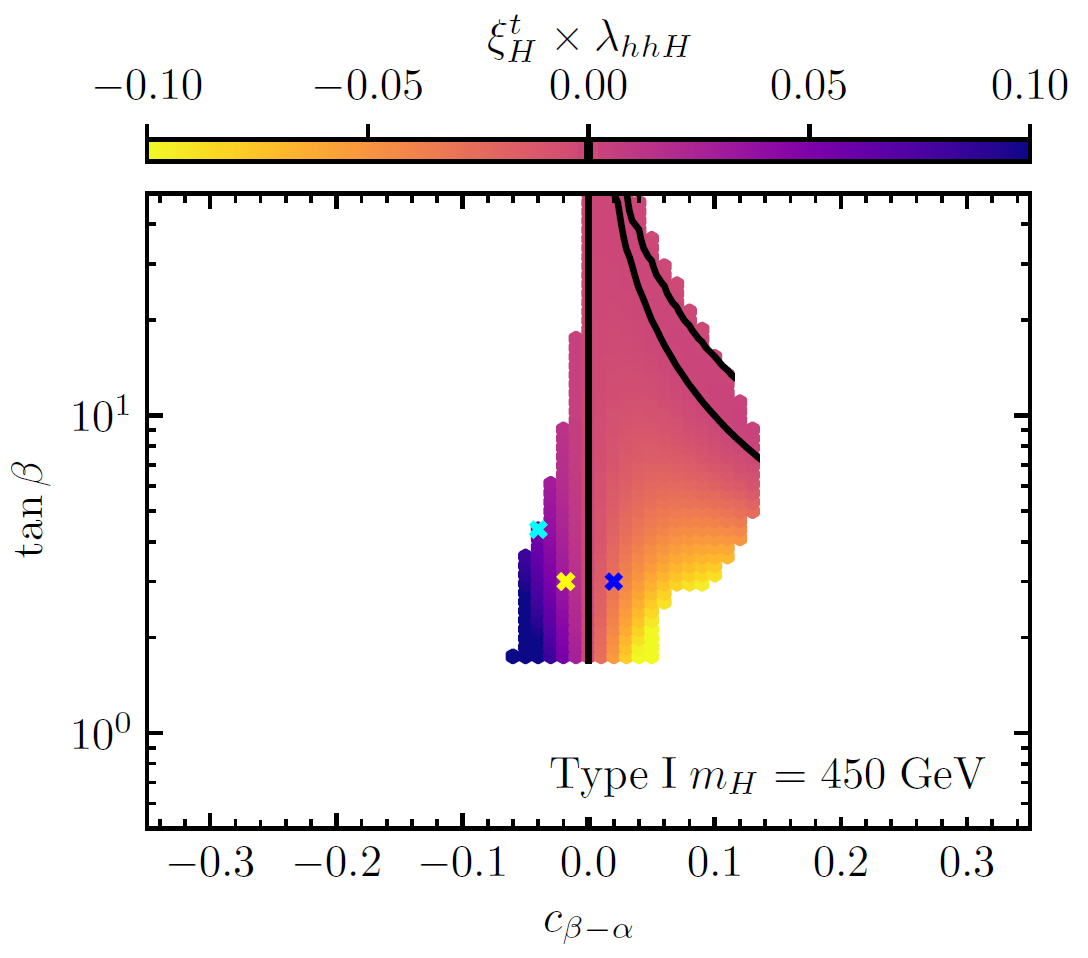}
\caption{\xila\ in the example benchmark plane in the Type~I 2HDM with $m_{\phi} = 450 \gev$ 
and $\msq$ fixed via \protect\refeq{msq}. Black lines are located at $\xila$ = 0, either because $\lahhH$ =0 (in the alignment
limit, i.e.~the vertical line, and the lower curved line on the right upper corner) or because $\xi_H^t$ = 0 (the upper line on the right corner). The blue cross 
indicates an example point whose $\mhh$ distribution is displayed in \protect\reffi{fig:smearbin}. The yellow cross
features an example with roughly the opposite sign value of $\xila$ and its distribution is displayed in
\protect\reffi{fig:posneg}. The cyan cross is an example point for which different predictions of the NN will
be analyzed further in \protect\reffi{fig:ex_4b_1unc}.}
\label{fig:prod_mH450}
\end{figure}

In our study we will concentrate on one exemplary benchmark plane, which is defined in the 2HDM Type~I. 
In our analysis we will only include parameter points that are in agreement with the constraints discussed in the 
previous subsection.
The example is chosen to be representative of a scenario with resonant di-Higgs production: 
We have set $m_{\phi} = 450 \gev$, which is assumed to be measured in a different process, i.e.\ with a different,
less complex final state, where later the corresponding experimental uncertainty will be considered. We have taken $\TB$ and $\CBA$ as free parameters, assuming that their
determination in the future will be more complicated. $\msq$ is either fixed via\footnote{ This relation ensures that perturbative unitarity is fulfilled for large $\TB$.}
\begin{align}
    \msq \equiv \MH^2 c_{\al}^2/\TB~,
    \label{msq}
\end{align}
or taken as a free parameter (see below). For this parameter the experimental prospects at the HL-LHC
are unclear~\cite{Arco:2022jrt}, and a variation of $\msq$ as a third free parameter is the most conservative approach.
In \reffi{fig:prod_mH450} we show our benchmark plane for $\msq$ fixed according to \refeq{msq}. 
The colored region is allowed by all the applied constraints while the white region is excluded by either of them. In particular, the lower bound on $\TB$ is set by flavor constraints, the bound on negative $\CBA$ comes from requiring the stability of the vacuum at tree level and for positive $\CBA$ and large $\TB$ from requiring perturbative unitarity at tree level. Additionally, for lower $\TB$ values and larger $\CBA$, searches for resonant di-Higgs production provide the most stringent constraints. For intermediate values of $\CBA$ the most stringent bound comes from the requirement of compatibility of the Higgs boson at 125 GeV signal strengths with the experimental measurements.
The color coding indicates the product of the couplings \xila\ that we will later aim to extract.
The black lines denote, where we find $\xila = 0$, and the three crosses mark benchmark points discussed below.
It can be observed that $\xila = 0$ for $\CBA = 0$, i.e.\ in the alignment limit, as expected. However, the value of 
$\xila = 0$ is also found along a curve for $\lahhH$ = 0 at $\CBA = 0.02-0.12$ and $\TB = 8-50$, 
as well as for $\xi_H^t = 0$ at $\CBA = 0.03-0.10$ and $\TB = 15-50$.
The values for \xila\ (in various set-ups) together with the corresponding \mhh\ distributions are going to be 
the input for a neural network as will be described below.

\subsection{Invariant Mass Distributions}
\label{sec:mhh}

The core of our analysis will be the connection between the value of \xila\ and the shapes of the corresponding
\mhh\ distributions.
The shape of the invariant mass distribution of di-Higgs production has been analyzed with great detail in the past.
In \citere{Capozi:2019xsi} even the possibility of classifying different kinds of distributions through a neural network
approach was investigated. The projected shapes were classified in the region of the coupling parameters space in an EFT 
approach, allowing for the identification of deviations in the couplings involved in the SM gluon fusion di-Higgs production 
trough the invariant mass shape analysis. However, since no further BSM state was
assumed, there was no investigation of the role of a BSM THC.

In previous works we have demonstrated that di-Higgs invariant mass distributions at the HL-LHC could possibly provide access to 
the BSM THC $\lahhH$~\cite{Arco:2022lai}.%
\footnote{Corresponding analyses for future $e^+e^-$ colliders can be found in \citeres{Arco:2021bvf,Arco:2025nii,Arco:2025pgx}.}%
~In particular, assuming that a resonant scenario is realized, i.e.\ the
contribution of $H$ is sizable,  
the sign of the product \xila\ would determine the resonant structure at $\mhh \approx \MH$. 
More precisely, the structure would be dip-peak for an overall negative sign and peak-dip for an overall positive sign,
assuming that no further BSM effects arise from e.g. loop corrections to the $\lahhh$ and $\lahhH$ trilinear Higgs coupling
(see \citeres{Arco:2022lai,Heinemeyer:2024hxa} for further details). However, these effects are partially washed out by experimental
uncertainties, as we briefly review in the next subsection.


\subsection{Experimental Challenges}
\label{sec:exp-chall}

\begin{figure}[t!]
\centering
\includegraphics[width=0.45\textwidth]{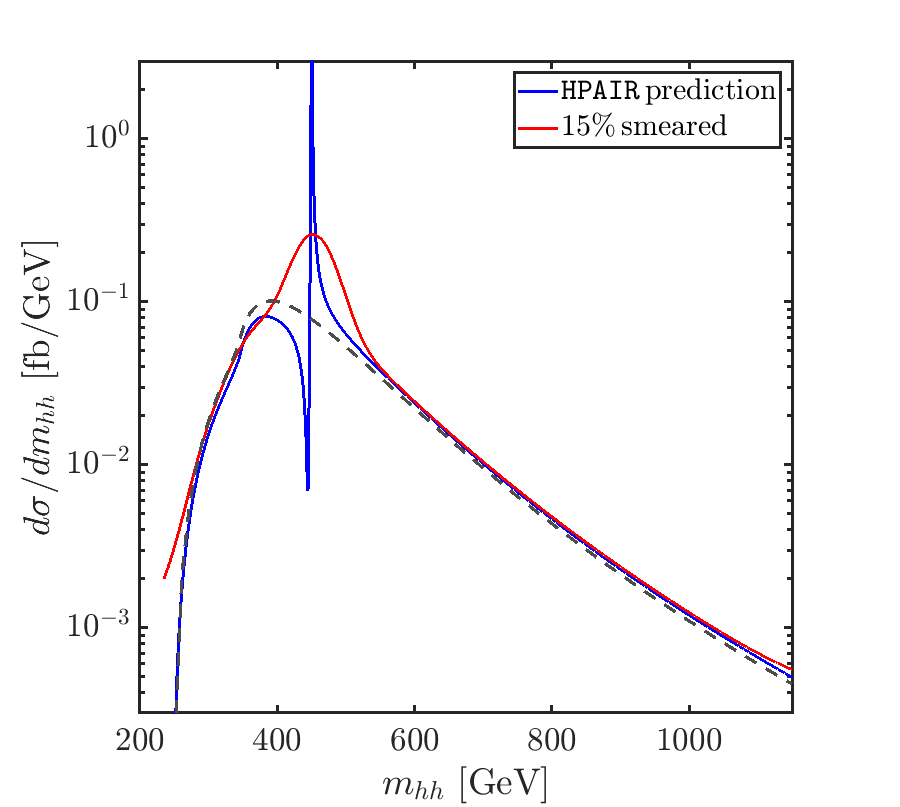}
\includegraphics[width=0.47\textwidth]{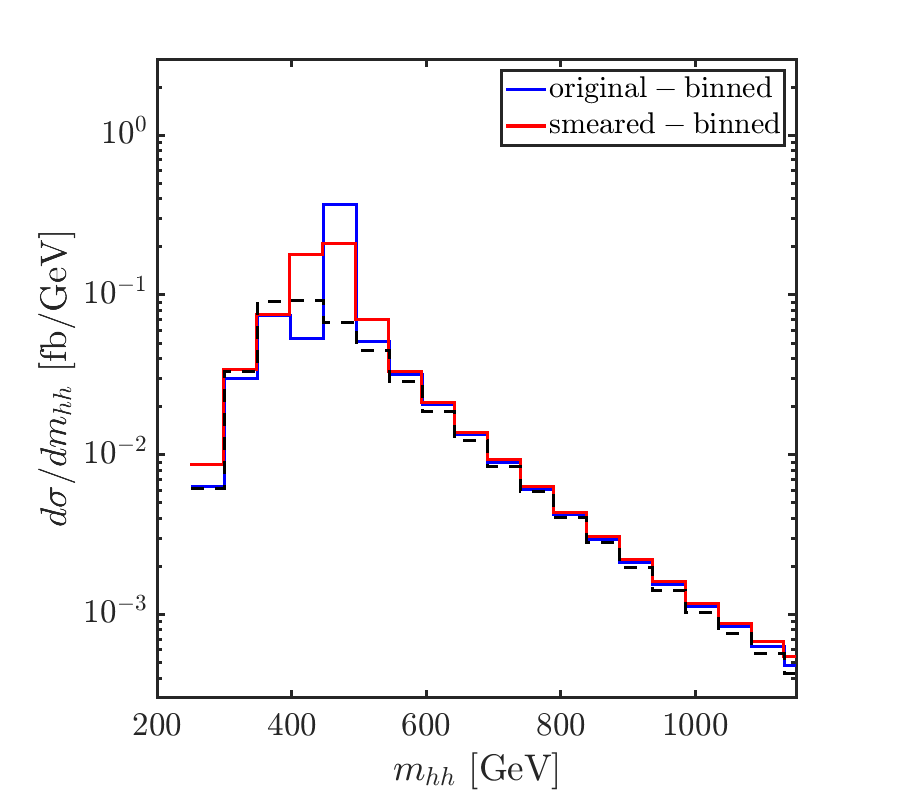}
\caption{Smearing (left) and binning (right) applied to the invariant mass distribution prediction for an example benchmark point in the 2HDM, marked in blue in \reffi{fig:prod_mH450}. The dashed lines indicate the continuum contribution.}
\label{fig:smearbin}
\end{figure}

An example for the theory prediction for the \mhh\ distribution, which is the key quantity for our analysis,
is shown in the left plot of \reffi{fig:smearbin}. The parameter point is taken from \reffi{fig:prod_mH450},
marked there with a blue cross, with $\CBA = 0.02$ and $\TB = 3$. 
For this point, we find a value of $\xila = -0.029$. The solid blue line shows the 
full \mhh\ distribution, while the dashed black line is calculated from the two continuum diagrams only, i.e.\ 
leaving out the $H$-resonance contribution. The dip-peak structure at $\mhh = \MH = 450 \gev$ is clearly visible.

However, the theory prediction is subject to experimental uncertainties, see \citere{Arco:2022lai} for a detailed
analysis, used to obtain our anticipated uncertainties. 
The first takes into account the uncertainty in the \mhh\ measurement. Each point of the
\mhh\ distribution is ``smeared" with a Gauss distribution with a width of 15\%, as shown in the red curves 
in \reffi{fig:smearbin}. It can be observed in the solid red line in the left plot that the smearing 
significantly dilutes the dip--peak structure and only a very small broader dip, followed by a broad ``hump", remains.


\thispagestyle{empty}

\begin{figure}[ht!]
\centering
\includegraphics[width=0.45\textwidth]{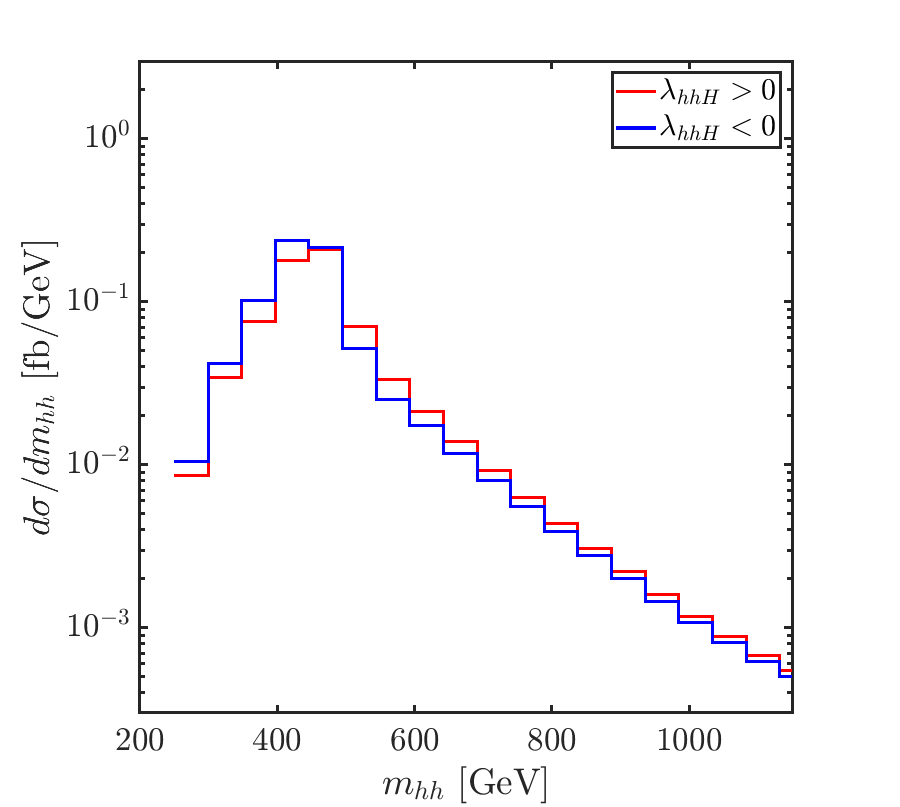}
\caption{Comparison of two benchmark points yielding a positive (blue) and negative (red) value of $\xila$
with $\xila \approx \pm 0.03$. The blue (red) curve corresponds to the point marked by a blue (yellow) cross in
\reffi{fig:prod_mH450}.} 
\label{fig:posneg}
\end{figure}

On top of that, experimental data is gathered in bins, owing to the finite energy resolution,
which may hide some of the effects that would be visible if the resolution was perfect. The bin size is set
to $50 \gev$, resulting in 16~bins (which serve later as input for our NN analysis).
In the right plot of \reffi{fig:smearbin} the solid blue curve shows the binned distribution for the unsmeared blue curve 
of the left plot. The solid red curve is the binned distribution of the smeared red curve in the left plot. 
Both are compared to the smeared and binned continuum distribution, i.e.~the distribution neglecting the resonance,
shown as black dashed curve. It can be observed that in this case the dip--peak structure around the resonance persists 
only very mildly (when compared to the SM prediction) after both smearing and binning have been taken into account.
If one further compares the red curve of \reffi{fig:smearbin} with a different one that yields a positive value of $\xila$, 
the peak--dip / dip--peak structure cannot be resolved optically anymore.
This is demonstrated in \reffi{fig:posneg}, where we show such a comparison, taking a point from \reffi{fig:prod_mH450} with 
$\TB = 3$, $\CBA = -0.018$. This yields a value of $\xila$ = 0.0304 (this point is marked with a yellow cross 
in \reffi{fig:prod_mH450}), which is roughly the same magnitude but opposite sign w.r.t. the one giving rise to the red line. 
This poses a challenge on the experimental access to $\xila$ that we will try to address in the following sections.

\thispagestyle{empty}


It should be noted that the numbers chosen for smearing  (15\%) and binning ($50 \gev$) were 
recommended within the LHC Higgs working group and are expected to
correspond to a reasonable (realistic) experimental set up.

\subsection{Statistical Uncertainties}
\label{sec:stat_unc}

The final location of the data points of an experimental distribution will also be affected by an unavoidable 
uncertainty of the data itself. This consists of three components, a statistical uncertainty, a theoretical uncertainty
(from unknown higher-order corrections) 
and a systematic uncertainty. Here we will only consider the first one because the latter two are harder
to estimate and in an optimistic scenario should be subdominant.
In order to obtain the statistical uncertainty we have to determine the expected number of events in each 
bin of the $\mhh$ distribution. Experimental efficiencies are only available for definite final states, i.e.\ 
taking into account the decay of the $hh$ system. A combination of various final states, on the other hand, is an 
experimental analysis on its own and thus goes far beyond the scope of our NN focused work.
The most promising channel in this regard (i.e.~the one producing the largest number of events taking into
account experimental efficiencies) is $gg \to hh \rightarrow b\bar{b}\,b\bar{b}$. 
We have calculated the expected number of events in this channel in each bin as:
\begin{equation}
N_i \; = \; \sigma_i(gg \rightarrow hh)\, \; \times \; \mathcal{L} \; \times \; \br^2(h \to b\bar{b}) \; \times \; \epsilon,
    \label{eq:N}
\end{equation}
where $\sigma_i(gg \to hh)$ is the differential production cross section in each $\mhh$ bin~$i$
times the size of the bin (in our case $50 \gev$), 
$\mathcal{L} = 3000 \ifb$ is the integrated luminosity expected at the end of the HL-LHC, 
$\br(h \to b\bar{b}) = 0.5841$ is the branching ratio of the decay of a SM Higgs boson
into a pair of bottom quarks. 
Finally, $\epsilon \equiv \epsilon_{\rm TOT}\,\epsilon_{\rm SR}$ is the detector efficiency. Here 
$\epsilon_{\rm TOT}$ is the preselection efficiency, in this case it is the number of events with $\leq$ 2 $b$-tagged 
jets over the total number of events ($N = \sigma \, \times \, \br \, \times \, \mathcal{L}$), 
and $ \epsilon_{\rm SR}$ is the efficiency of the signal region (SR), i.e.\ the number of di-Higgs events
out of the preselected events. In practice, $N_i$, the number of events in each bin, is expected to follow a Poisson distribution with the mean given by~\refeq{eq:N}.
For the $b\bar b\, b \bar b$ channel we took the efficiencies 
from Fig.~3 (right plot for $s=0$) in \citere{ATLAS:2022hwc} for $\MH = 450 \gev$.
They are 17.3\% and 1\% for the total and the signal region efficiency, respectively.
The statistical uncertainty in bin~$i$ is then given by $\sqrt{N_i}$,
where $N_i$ is given by \refeq{eq:N}.

\begin{figure}[ht!]
\centering
\includegraphics[width=0.54\textwidth]{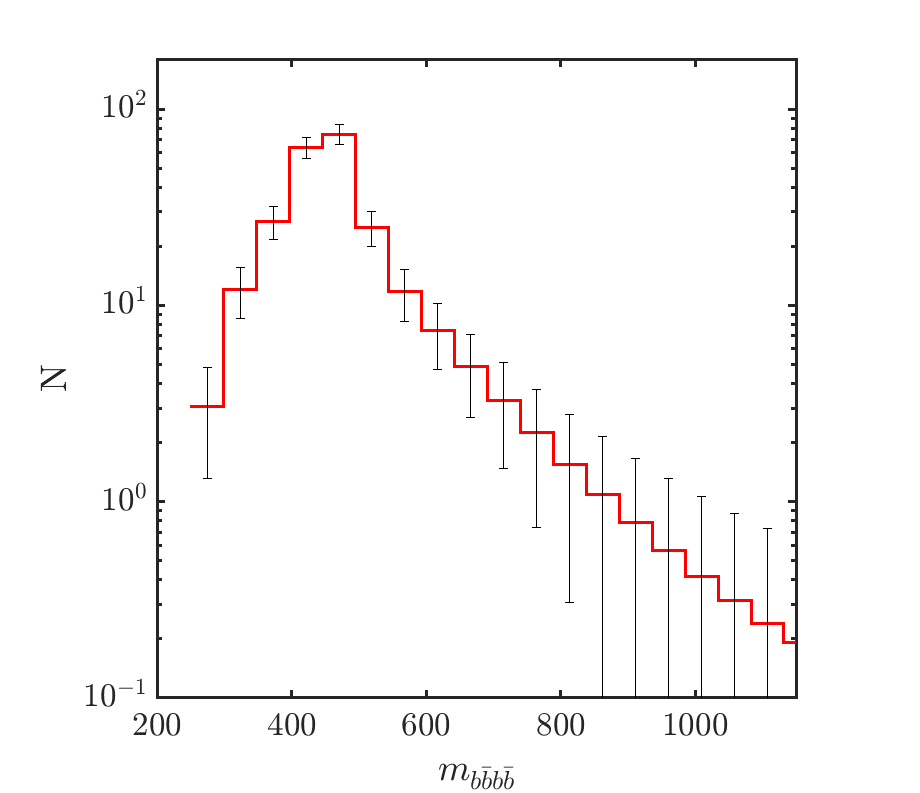}
\caption{Example number of events in the $b\bar{b}\,b\bar{b}$ channel with statistical error bars (see text)
for the point marked by the blue cross in \reffi{fig:prod_mH450}.}
\label{fig:atlas4bexample}
\end{figure}

In \reffi{fig:atlas4bexample} we show one example of the number of events in the $b\bar{b}\,b\bar{b}$ channel for 
the parameter point of the benchmark scenario with $\MH = 450 \gev$ marked by the blue cross 
in \reffi{fig:prod_mH450}. The statistical
uncertainties are shown as black error bars in each $\mhh$ bin. These statistical
uncertainties will be interpreted as one standard deviation of a Poisson distribution around the (theoretical) true value, which is the mean of the Poisson distribution.


\section{Statistical Analysis}
\label{sec:stat}

In this section we investigate the determination of \xila\ with classical methods, i.e.\ not using NN. Statistically, we can perform an hypothesis
test of the SM. We compare it to the null hypothesis ($H_0$), which is $\xila = 0$ (as realized in the SM), to an alternative hypothesis ($H_1$),
which is the full 2HDM prediction including the $H$-resonance contribution.
The deviation is then evaluated for our parameter of interest, \xila. It should be kept in mind that
we take the 2HDM just as a plausible model where such non-zero values of the BSM trilinear couplings can be accommodated.
Our results should be considered more general, however, i.e.\ a similar analysis can be performed in any model containing a heavy $\cp$-even
Higgs resonance contributing to di-Higgs production at the HL-LHC.
For the aforementioned hypothesis test we compute the likelihood ratio test between the null model (which gives rise to no resonance 
and \xila $= 0$), and the alternative model (where \xila $\neq 0$).

As the observed values in each bin are modeled as samples from a Poisson distribution, the likelihood of each hypothesis is given by
\begin{equation}
    L(H_0) = \prod_{i=0}^{15} \frac{{m_i}^{n_i} e^{-m_i}}{{n_i}!}\; \mbox{~and~}\;
    L(H_1) = \prod_{i=0}^{15} \frac{{n_i}^{n_i} e^{-n_i}}{{n_i}!},
    \label{eq:likelihood}
\end{equation}

\noindent where $m_i$ is the predicted mean count of the number of events in each bin in the 2HDM, and $n_i$ is the observed count from experimental data.
The likelihood ratio (LR) test statistic, $\lambda_{\mathrm{LR}}$, is defined as
\begin{equation}
    \lambda_{\mathrm{LR}} = -2 \ln \left( \frac{L(H_0)}{L(H_1)} \right) = 2 \sum_{i=0}^{N_{\mathrm{bins}}} \left (n_i \ln \frac{n_i}{m_i} + m_i - n_i \right).
\end{equation}
According to Wilk's theorem, $\lambda_{\mathrm{LR}}$ asymptotically follows a $\chi^2$ distribution, provided that the sample size is large enough: As a rule of thumb each $n_i$ and $m_i$ should be larger than 5. This theorem allows to directly compute a $p$-value as 
\begin{equation}
    p = P(\chi^2>\lambda_{\mathrm{LR}})\,.
\end{equation}
The $p$-value is the probability ($P$) of obtaining a result at least as extreme as the data, assuming the null hypothesis is true. Typically values 
of $p<0.05$ (roughly corresponding to 95\% CL) suggest evidence against the null hypothesis. In order to correctly apply Wilk's theorem, we need 
to ensure that the number of events in each bin is ``large enough". Therefore we aggregate some bins together in order to have better statistics. We 
show an example of this aggregation in \reffi{fig:aggregated_bins}, where two sample distributions from \reffi{fig:prod_mH450} are shown:
one with a small product of couplings,  $\xila = 0.0188$ (orange), and one with a large value, $\xila = 0.0553$ (green). They are
compared to the SM distribution (blue), which is equivalent to the alignment limit. 
On the left we show the distributions with the original 50 GeV binning and on the right the aggregated bins that lead to a particle count of at 
least 4 events, so that the statistical tests can be performed.\footnote{We have performed a statistical analysis only using the bins which initially have $n_i\geq5$ events. The result for that case is similar as the one using aggregated bins.}
In particular, the first and second bin are added up together, and correspond to the first bin on the right. The third, fourth, fifth and sixth bin 
are unchanged and correspond to the bins 2, 3, 4 and 5 on the right. All the other bins are summed in bin number 6 on the right. We use the aggregated bins for the statistical analysis performed below, for the NN we keep the extended 16-bin sample, as the statistical considerations discussed above do not play a role in the NN.

\begin{figure}[ht!]
\centering
\includegraphics[width=0.9\textwidth]{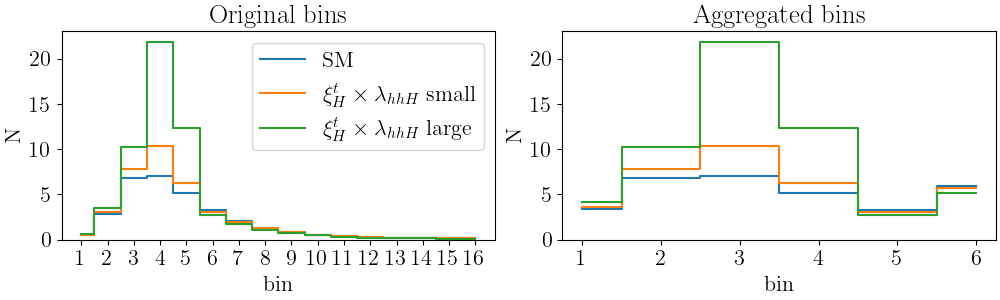}
\caption{Sample distributions for the number of events with the original 50 GeV binning (left) and the aggregated bins (right). Shown are the distributions for the SM (blue), and two examples with a small value of $\xila = 0.0188$ (orange), and with a large value,
$\xila = 0.0553$ (green).} 
\label{fig:aggregated_bins}
\end{figure}
\thispagestyle{empty}

The $p$-values for the benchmark plane shown in \reffi{fig:prod_mH450} are depicted in \reffi{fig:pvalue}. For the parameter points for which the computed
$p$-values are less than 0.05 (colored in blue), the null hypothesis can be rejected. This method is not used for parameter estimation, but ``just'' for hypothesis
testing, and the bottom line is that the distributions of the points within the red regions are compatible with the SM at 95\% CL.

\begin{figure}[ht!]
\centering
\includegraphics[width=0.45\textwidth]{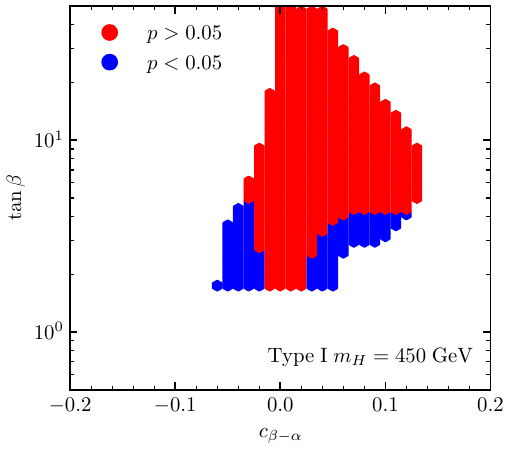}
\caption{$p$-value in the example benchmark plane. The red regions shows the parameters points where the p-value is larger than 0.05, i.e.~the null hypothesis cannot be rejected. The blue colored points represent the region where the null hypothesis can be tested with classical methods and a deviation from $\xila = 0$ is statistically significant.}
\label{fig:pvalue}
\end{figure}
\thispagestyle{empty}

As a side note, the square root of the a $\chi^2$-distributed variable, in our case $\lambda_{\mathrm{LR}}$, can be interpreted in terms of standard deviations from the null hypothesis. In particular if  $\sqrt{\lambda_{\mathrm{LR}}}>1.96$, the conclusion would be the same as if $p<0.05$, meaning such a point is statistically far from the null hypothesis. However, this approach only works exactly for~1 degree of freedom, and gets worse for our case, where we have 6 degrees of freedom (i.e.~the $N_{\mathrm{bins}}=6$ after aggregation). For this reason, we show in \reffi{fig:pvalue} the $p$-values for the benchmark plane
defined in \reffi{fig:prod_mH450}, rather than the significance.

The $p$-value is a method used for hypothesis testing, and cannot really serve the purpose of our problem. We want to estimate a parameter \xila\, from a distribution. Classically, for parameter estimation, the most precise method is the maximum likelihood estimation (MLE). The likelihood of each BSM scenario is defined as in \refeq{eq:likelihood}. We calculate $L (H_0)$ for each set of $m_i$ (all points in \reffi{fig:prod_mH450}) and find the one that maximizes $L$ given the experimentally observed count $n_i$. This would be the most likely parameter point to produce the observed distribution. This method is only meaningful if the input data is assumed to be Poisson distributed around the true value. 
Otherwise the theoretically predicted point will trivially be the best fit. 

\begin{figure}[ht!]
\centering
\includegraphics[width=0.45\textwidth]{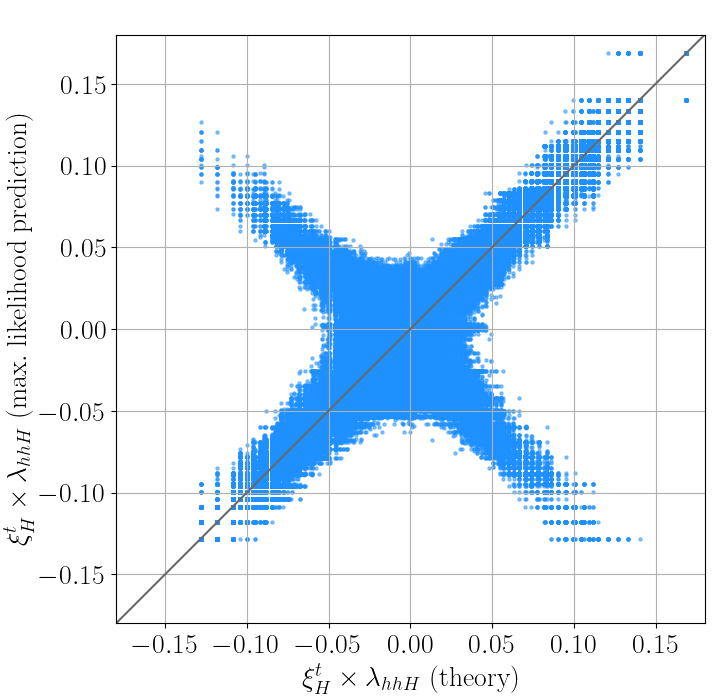}
\caption{Predicted value of $\xila$ with the classical maximum likelihood estimation method versus the theoretical value of $\xila$.}
\label{fig:maxlikelihoodest}
\end{figure}
\thispagestyle{empty}
%
Figure~\ref{fig:maxlikelihoodest} shows the predicted parameter \xila\ with the MLE method versus the theoretical \xila. In total, 
our sample consists of 2048 \mhh\ distributions, smeared according to their Poisson distribution. 
We see that the method fails to correctly predict the sign of the \xila\ couplings, as we expected from the invariant mass distributions. The 'X' shape in~\reffi{fig:maxlikelihoodest} explicitly demonstrates that while the method manages to determine if $\xila \neq 0$,  the sign of $\xila$ cannot be predicted. These limitations necessitate the exploration of other techniques, such as neural networks, to achieve the required sensitivity and precision for the numerical estimation of $\xila$.\footnote{Classical methods can still be competitive in the determination of other BSM couplings such as the $\CP$-violating top Yukawa interaction~\cite{Bahl:2024tjy}.}

\section{Determination of \boldmath{\xila} with NNs}
\label{sec:method}

In this work, we focus on the following problem: the extraction of the value of $\xila$
from the $\mhh$ distribution, as anticipated to be measured at the HL-LHC. 
We employ a neural network that has been trained on ``realistic" \mhh\ distributions, i.e.\ taking
into account smearing, binning, and, most crucially, Poisson noise to reflect statistical fluctuations in future measurements. This approach allows us to simulate the expected conditions under which experimental data would be obtained, making our analysis as close to reality as possible.

We employ a simple one layer network using 
\texttt{PyTorch}~\cite{paszke2019pytorchimperativestylehighperformance}. The network architecture consists of an initial batch normalization layer, followed by a single hidden layer with 64 neurons and ReLU activation, and a final output neuron with a linear activation to predict the value of \xila. We use the Adam optimizer (learning rate $10^{-4}$, no weight decay or regularization) with mean squared error (MSE) loss. It is important to emphasize that this neural network is almost trivially small. The key point here is that this is not an application of deep learning and associated concepts, but instead a demonstration that a universal function approximator like a neural network can be used to replace classical statistical methods. In our case, we show that the network can learn to predict a value from simulated experimental data with a precision exceeding that of classical methods. More complex architectures were tested but offered no significant performance advantage, despite requiring substantially more computational resources.

As input, the model receives 16-bin histograms of the smeared $\mhh$ distribution, with each bin representing a width of 50~GeV. 
First, these $\mhh$ distributions as obtained from the theory calculation are used for training. In a later stage, 
these inputs are modified by applying Poisson noise, mimicking statistical uncertainties on the binned experimental data. The batch normalization layer is used to automatically rescale the count values in each bin to values close to 0 in order to speed up convergence in training.
The training dataset consists of $\mathcal{O}(500 - 4000)$ distributions, depending on the scan settings of the 2HDM parameter space 
(i.e.~whether we fix $\msq$ and $m_H$ or not). 25\% of the data is optionally retained for validation, depending on the configuration 
(we retain it only if we do not apply Poisson noise for training). 
Each model is trained for a total of $2^{15}$ epochs, during which a new noisy version of the input is generated at each epoch.
In \reffi{fig:workflow} we attach a schematic diagram of the workflow, specifying the input and output in each step for the different codes that were used, as well as a diagram of the NN architecture.

\begin{figure}[ht!]
\centering
\includegraphics[width=0.69\textwidth]{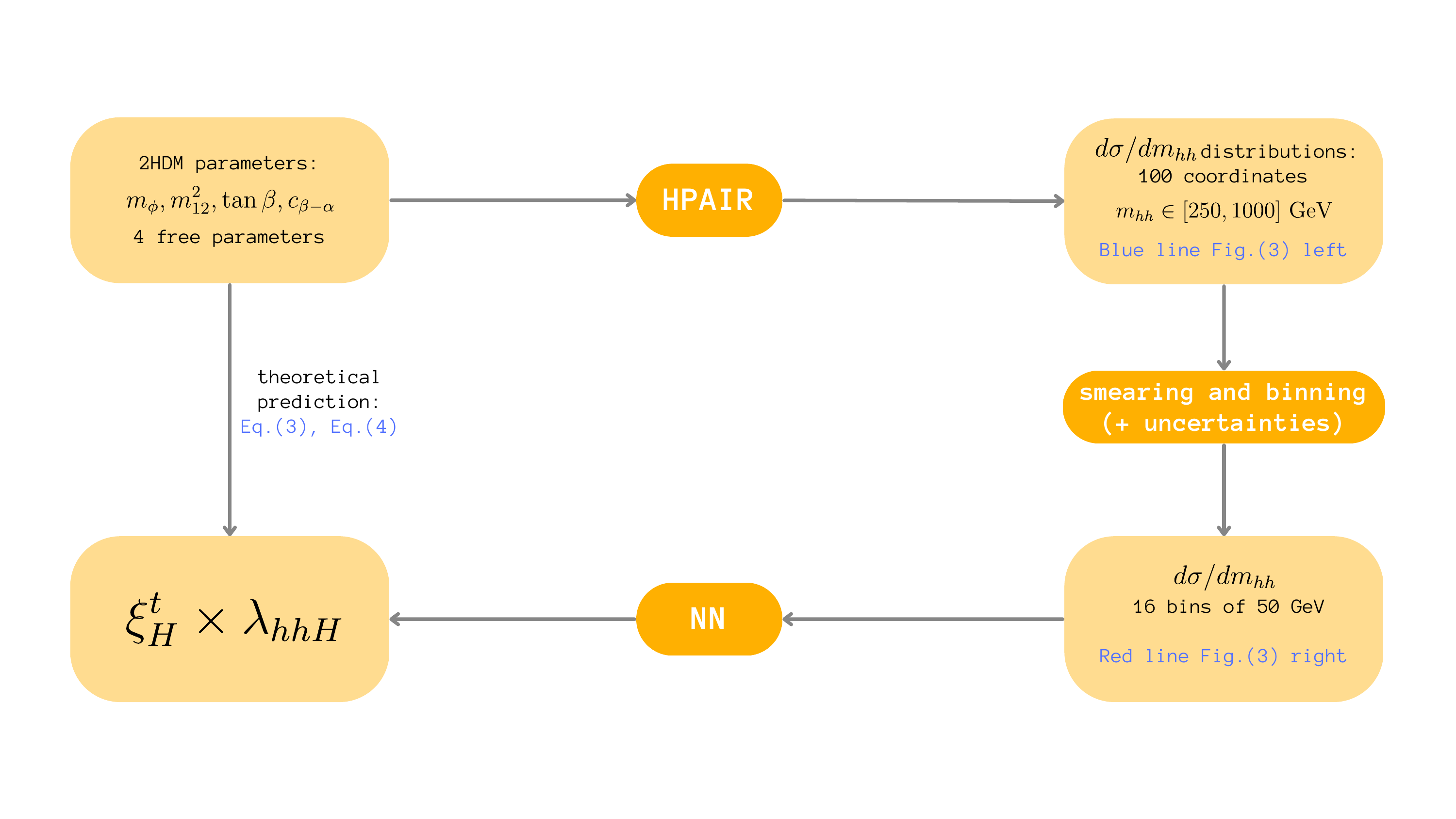}
\includegraphics[width=0.2\textwidth]{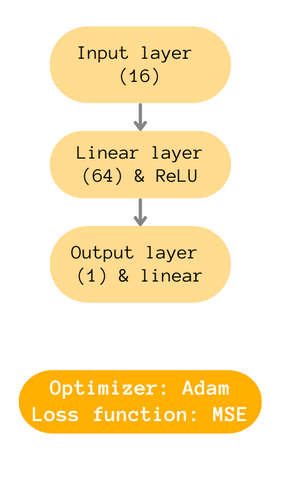}
\caption{Workflow (left) and NN architecture (right) (see text for details).}
\label{fig:workflow}
\end{figure}
\thispagestyle{empty}

It is worth to note that in
our approach, the NN has been trained completely on
parameter points of the 2HDM. However, we expect similar results from models where the di-Higgs production process 
is resonantly enhanced by the contribution of an additional heavy scalar. 
On the other hand, during the HL run of the LHC, the searches for BSM Higgs bosons will continue. We assume here 
a potentially realistic situation, in which the heavy $\cp$-even Higgs will have been discovered and its mass will have 
been measured with a sufficiently high accuracy (see below). Furthermore, the rest of the spectrum, the $\cp$-odd Higgs 
boson, as well as the pair of charged Higgs bosons will have been found (and their masses will have been determined, which
play no role in our analysis, but could influence the theoretical prediction of $\mhh$ around the resonance at higher orders).
In the case that in the future a different scalar sector will be detected experimentally, the 
NN analysis performed here will have to be adopted to the then favored Higgs-boson sector. While all this at the current
time is a hypothetical scenario, we use this as a viable example to demonstrate the extraction of \xila\ from the
di-Higgs boson measurements at the HL-LHC.



\section{Results}
\label{sec:results}

In this section, we will present the results of our analysis, starting from relatively simple scenarios to
more and more realistic ones by taking into account the various anticipated experimental uncertainties.
We train the NN and obtain the NN predictions as described in \refse{sec:method}, if not specified differently.


\subsection{Variation of \boldmath{$\msq$} vs.\ Fixed \boldmath{$\msq$}}

We start our analysis with the effects of a variation of $\msq$,
which is difficult to measure experimentally, see \citere{Arco:2022jrt} for some proposal.

In \reffi{fig:scan} we show the parameter scan of the benchmark scenario defined in \refse{sec:benchmark} projected in
two-dimensional planes of $\TB-\CBA$ (left), $\msq-\CBA$ (middle) and $\msq-\TB$ (right). 
The upper row shows the projections with $m_{12}^2$ restricted to values that obey 
$\msq = \MH^2 \cos\alpha^2/\TB$ according to \refeq{msq}. This condition, however, was artificially introduced to facilitate finding a parameter space in agreement with the theoretical constraints. In order to demonstrate the effects
of this artificial choice, in the lower row of \reffi{fig:scan} we show the three projections including all $\msq$ values that
yield agreement with the theoretical constraints. This results in a truly 3-dimensional scan in the three remaining free 
parameters (once $m_{\phi}$ is fixed), i.e.~$\CBA$, $\TB$, and $m_{12}^2$. In the lower right panel one can observe that for large $\TB$ values the allowed region for $\msq$ is constrained to the one that fulfills \refeq{msq}, while for low $\TB$ the value of $\msq$ remains relatively unconstrained. The color coding indicates the values of 
$\xila$ and the white regions are disallowed by either of the considered constraints.

\begin{figure}[ht!]
\centering
\includegraphics[width=0.32\textwidth]{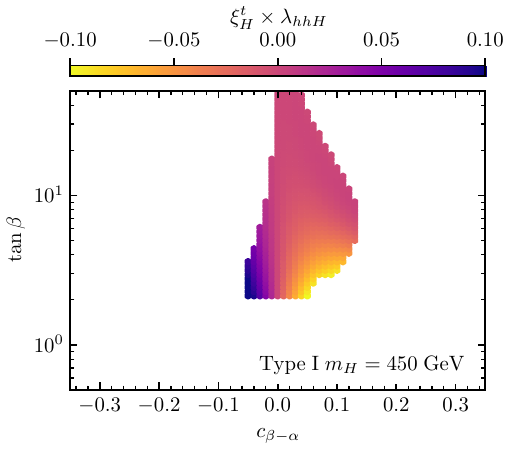}
\includegraphics[width=0.32\textwidth]{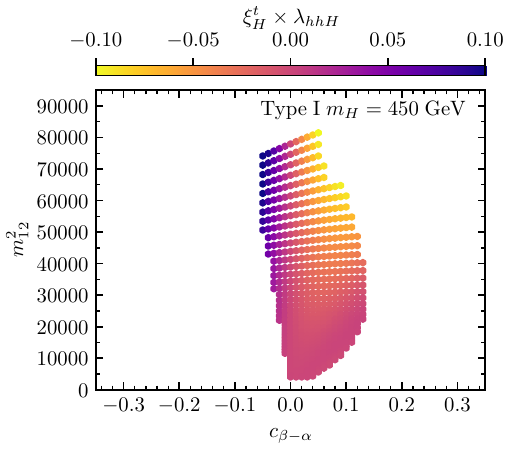}
\includegraphics[width=0.32\textwidth]{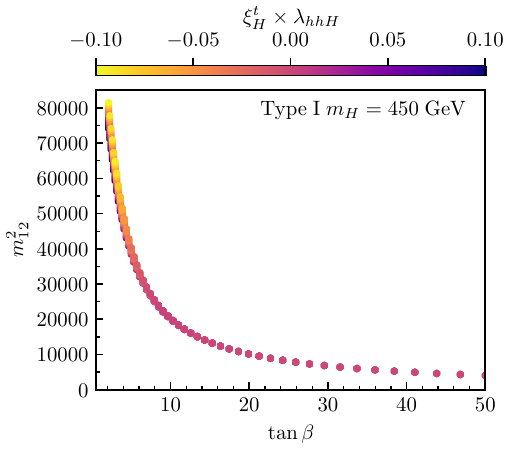}
\includegraphics[width=0.32\textwidth]{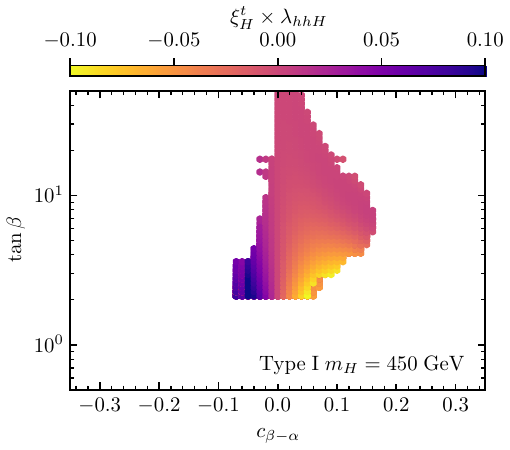}
\includegraphics[width=0.32\textwidth]{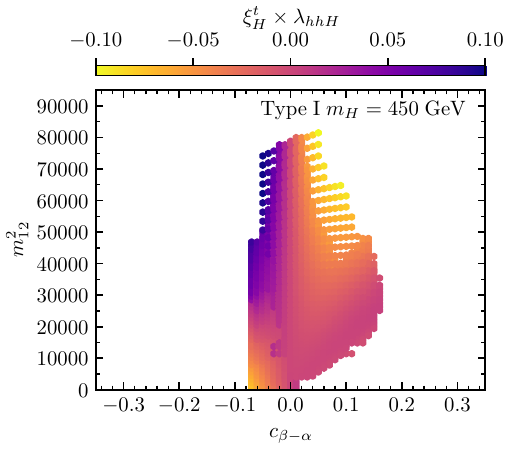}
\includegraphics[width=0.32\textwidth]{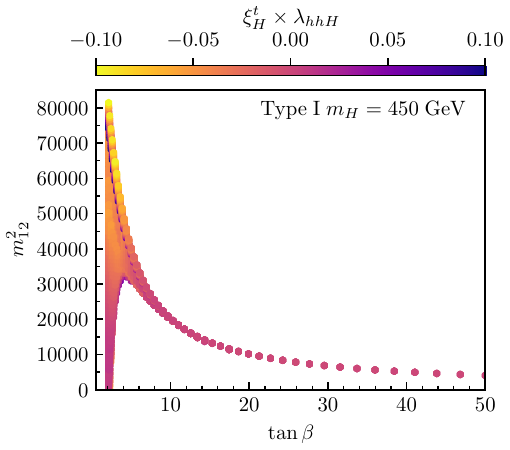}
\caption{Upper row: $\xila$ in the allowed region of the original benchmark plane with $\MH = \MHp = \MA = 450 \gev$ and 
$\msq = \MH^2 \cos\alpha^2/\TB$, projected in the three possible dimensions $\to$ planes.
Lower row: same of the above $\to$ Same as upper row, but for all theoretically allowed values of $m_{12}^2$ (see text).}
\label{fig:scan}
\end{figure}

We study the effect of not fixing the value of $m_{12}^2$ by applying the NN on the scenario with 
fixed $\msq$ in comparison with the case of varied $\msq$. In \reffi{fig:2dvs3d} we show the result of the prediction 
for $\xila$ when the model is trained on the benchmark plane fixing $\msq$ according to \refeq{msq} on the left
and without any assumption on the right. 
For these plots, we have trained the network on 75\% of the points of the corresponding scenario (fixed $\msq$
on the left and free $\msq$ on the right) and tested it on the other 25\% of the points. Using a 4-fold split of the full data (thus evaluating \xila\ for the whole allowed parameter space) 
we observe that the prediction of the parameter $\xila$ 
by the NN over the whole plane is perfect for both scenarios.
%
Thus, it
can be seen that the effect of varying $\msq$ does not have a relevant effect on our analysis. Consequently,
for the sake of computational efficiency we will continue the following analysis with the model based on the benchmark
plane described in Sec.~\ref{sec:benchmark}, i.e.\ with $\msq$ fixed according to \refeq{msq}.

\begin{figure}[ht!]
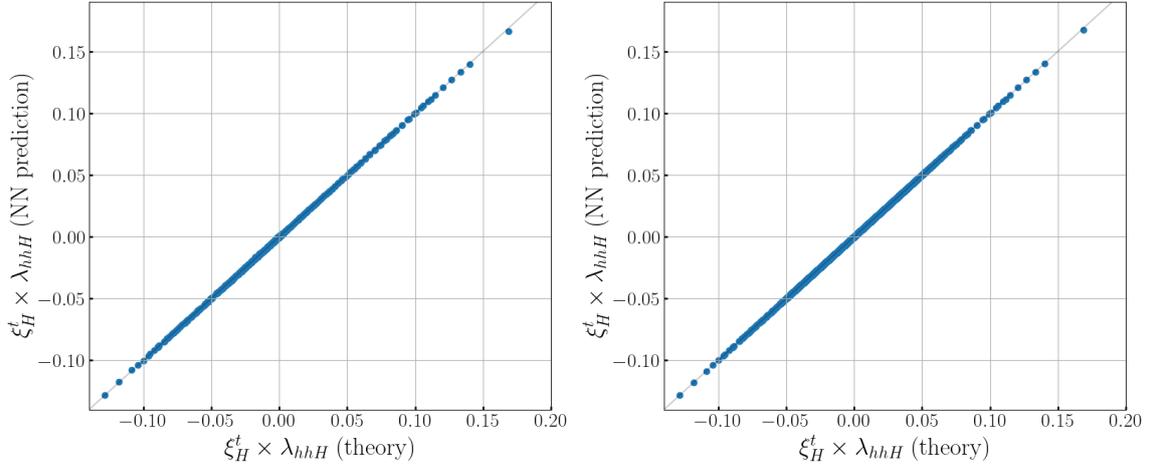

\centering
\includegraphics[width=0.45\textwidth]{figs/pred_450_2d.pdf}
\includegraphics[width=0.45\textwidth]{figs/pred_450_3d.pdf}
\caption{Prediction of $\xila$ in the original benchmark plane (left) and the three dimensional benchmark plane with $m_{\phi}=450 \gev$ and $\TB$, $\CBA$, and $m_{12}^2$ as free parameters (right).} 
\label{fig:2dvs3d}
\end{figure}


\subsection{Variation of \boldmath{$\MH$}}
\label{sec:NNmodels}

In order to investigate the NN results on the anticipated precision of $\MH$ we analyzed the results in 
four different ``models''. The NN is trained on
\begin{description}[font=\normalfont]
\item[\texttt{Model~I}:] only the example benchmark plane in \reffi{fig:prod_mH450} is used for the training.
\item[\texttt{Model~II}:] \texttt{Model~I} plus two planes with $\MH = 450 \pm 1 \gev$ and two planes with $\MH = 450 \pm 5 \gev$.
\item[\texttt{Model~III}:] \texttt{Model~II} plus two planes with $\MH = 450 \pm 10 \gev$.
\item[\texttt{Model~IV}:] \texttt{Model~III} plus two planes with $\MH = 450 \pm 15 \gev$.
\end{description}

\begin{figure}[ht!]
  \begin{center}
\includegraphics[width=0.45\textwidth]{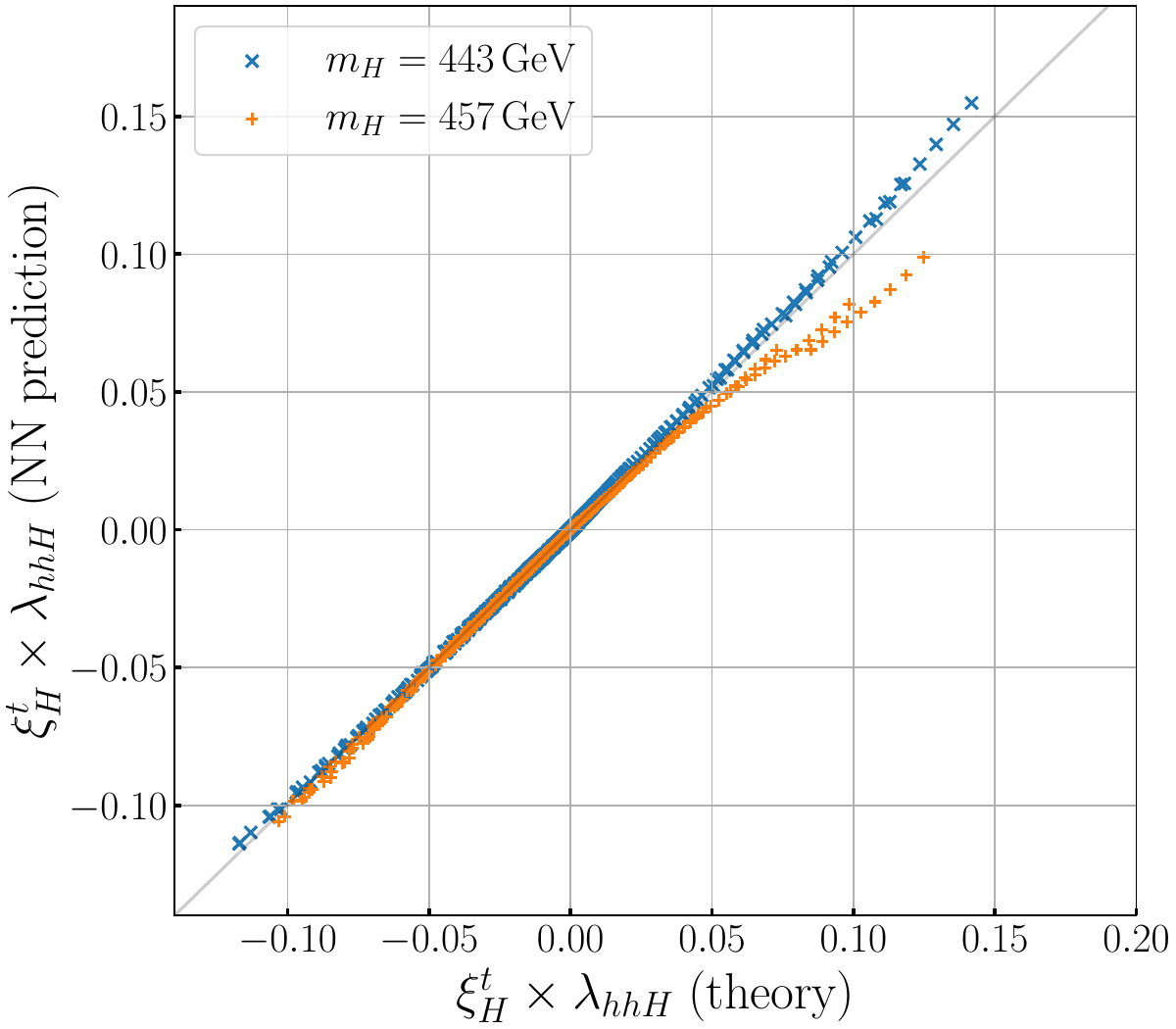}
\includegraphics[width=0.45\textwidth]{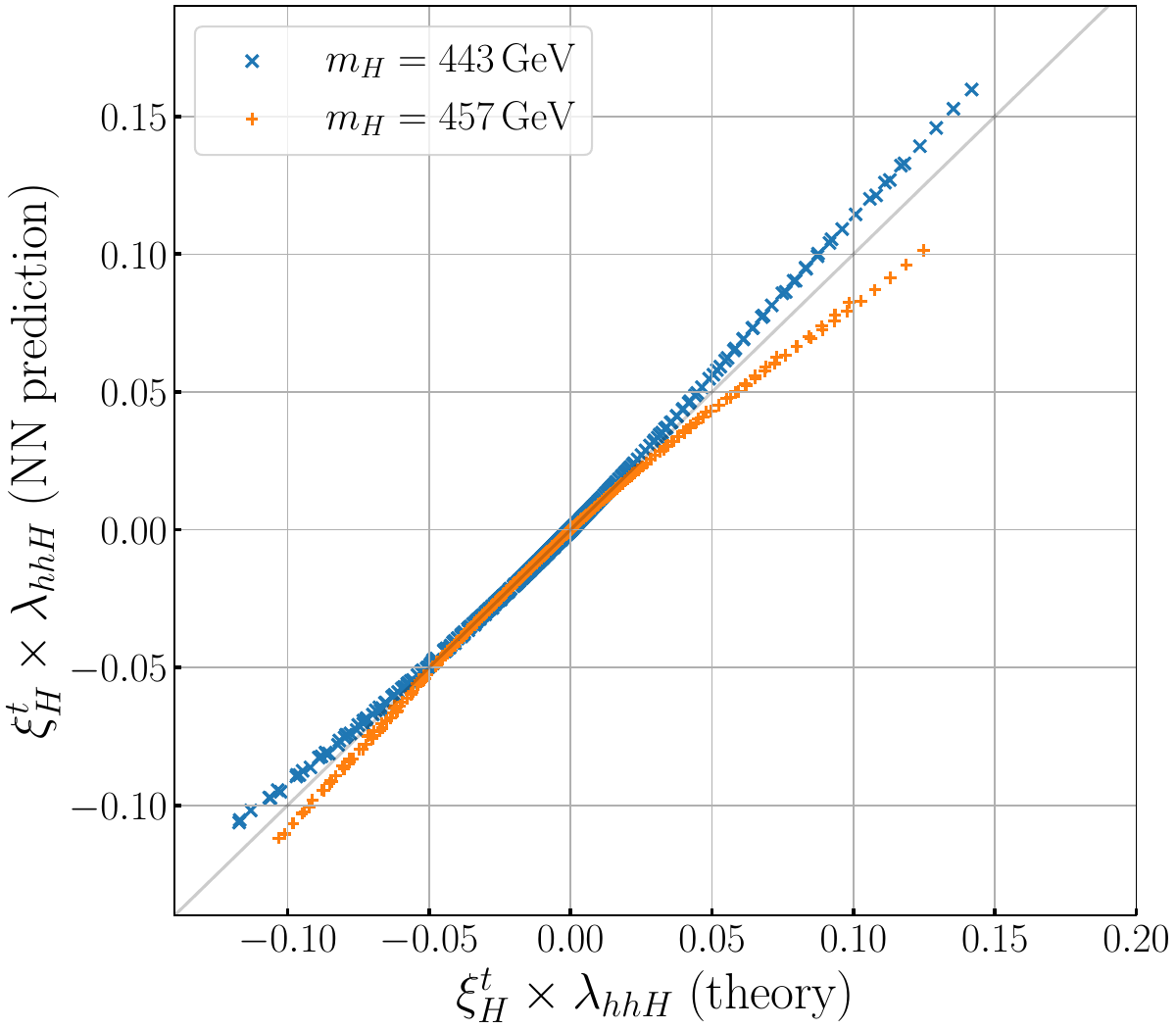}
\includegraphics[width=0.45\textwidth]{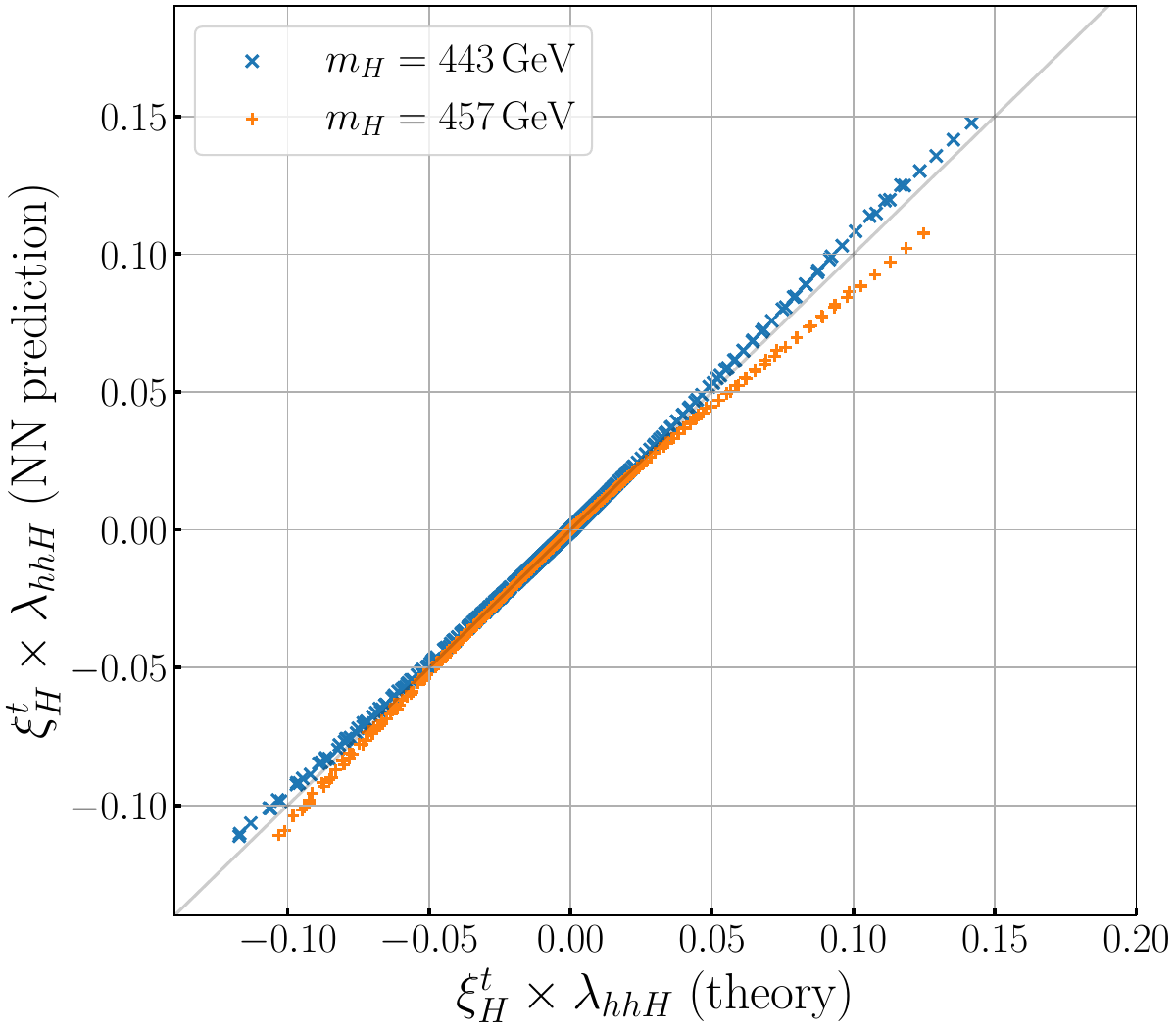}
\includegraphics[width=0.45\textwidth]{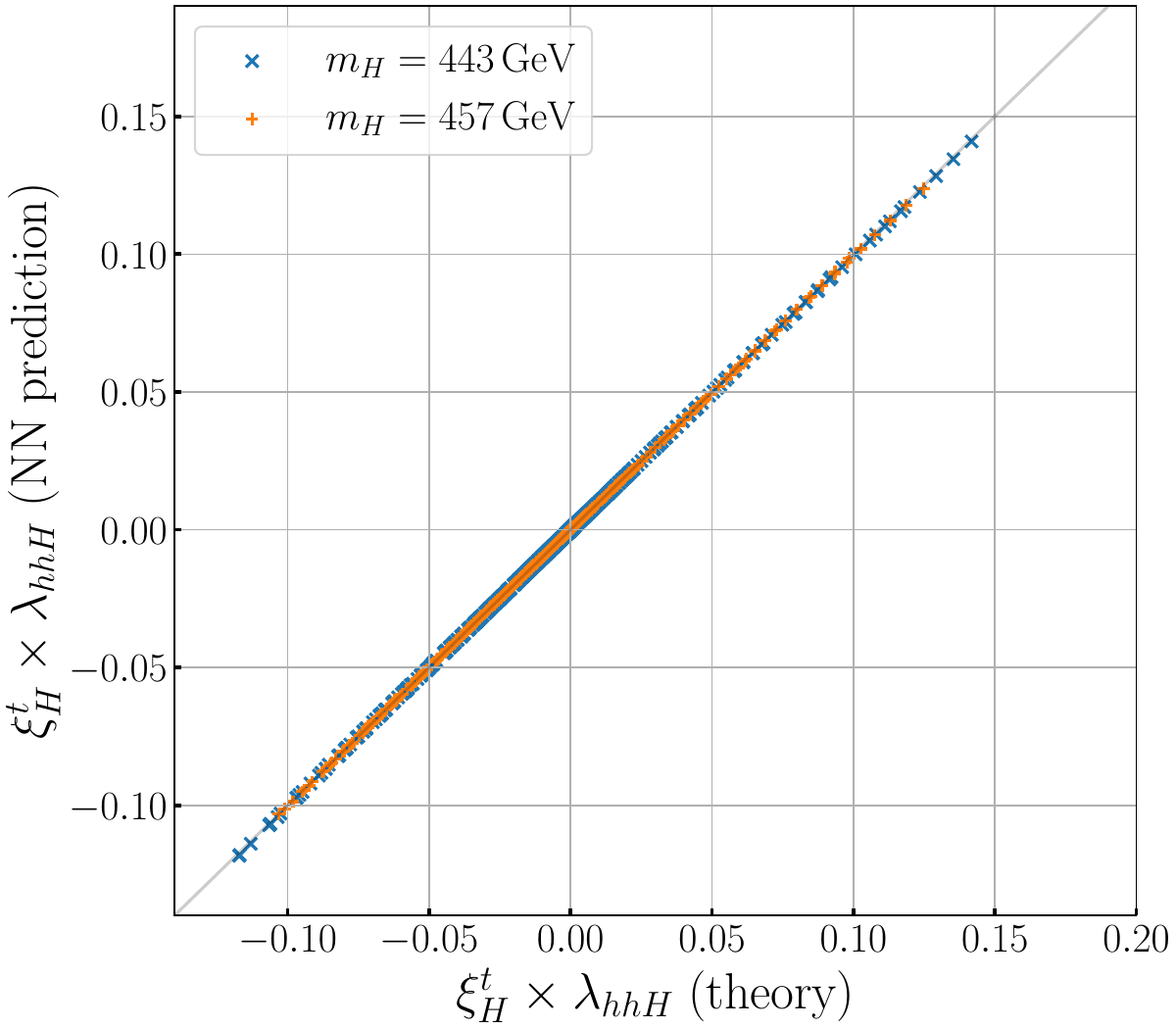}
	\end{center}
\caption{Results of our NN analysis for $\xi_H^t \times \lambda_{hhH}$ in \texttt{Model-I, -II, -III, -IV} in the upper left, upper right,
lower left and lower right plot, respectively. The blue (orange) points 
show the results for $\MH = 443 (457) \gev$ (see text). The gray line is there to guide the eye and corresponds to the coincidence of the values obtained from the NN with the values predicted by theory.}
\label{fig:test-mH}
\end{figure}

While in the previous subsection the nominal benchmark value of $\MH = 450 \gev$ has been used, we now assume
that the ``real'' value differs by $\pm 7 \gev$ from the measured one%
\footnote{The experimental precision on BSM Higgs-boson masses at the LHC remains to be analyzed in detail; 
see, however, \citere{Gennai:2007ys}.}%
, resulting in a corresponding distortion of the $\mhh$ distributions.
The results are shown in \reffi{fig:test-mH} for the four models defined above: 
the NN output of \texttt{Model-I, -II, -III, -IV} is presented in the upper left, upper right,
lower left and lower right plot, respectively. The blue (orange) points show the prediction of the NN for two distinct
test cases: $\MH = 443\,(457) \gev$.
In both these cases we show the values of $\xila$ predicted by the NN for the whole set of allowed parameter points 
in the $\CBA$--$\TB$ plane.
It can be observed
that the NN prediction improves significantly once we take into account as many datasets, i.e.\ a large 
variation of $\MH$, as possible in order to train the NN. We conclude that a possible uncertainty in $\MH$ can be 
brought easily under control by a sufficient NN training for different $\MH$ values, in particular we find that the dataset needs to contain points with an uncertainty twice as large as the measured values (e.g.~for a $\pm 7 $ GeV uncertainty in $m_H$, the dataset needs information from $\pm$ 15 GeV deviation in $m_H$ distributions). Consequently, in the following we will either keep $\MH$ fixed
to its nominal value of $\MH = 450 \gev$, or use \texttt{Model IV}, when the uncertainty on $m_H$ is taken into account.

\subsection{Experimental Uncertainties in the \boldmath{$\mhh$ Determination:\\ 
Final Results for the NN Determination of \boldmath{$\xila$}}}
\label{sec:resunc}

So far we have assumed that the experimental result for the measured $\mhh$ distribution agrees exactly with
the theoretical prediction. However, a realistic distribution is subject to statistical and systematic uncertainties
in the experimental measurement of $\mhh$, as discussed in Sec.~\ref{sec:stat_unc}. While an assessment of the systematic uncertainties goes far beyond the scope
of this work, for the final results we will include the anticipated statistical uncertainties in the $\mhh$ measurement. In order to take this into account, we repeat our analysis for the three cases discussed so far: 1) the initial benchmark scenario with $m_H$ and $m_{12}^2$ fixed, 2) relaxing the constraint on $m_{12}^2$, and 3) relaxing the constraint on $m_H$.%
\footnote{Relaxing both constraints at the same time would lead to a very large increase in the data points that would be computationally challenging.}

\begin{figure}[ht!]
\centering
\includegraphics[width=0.45\textwidth]{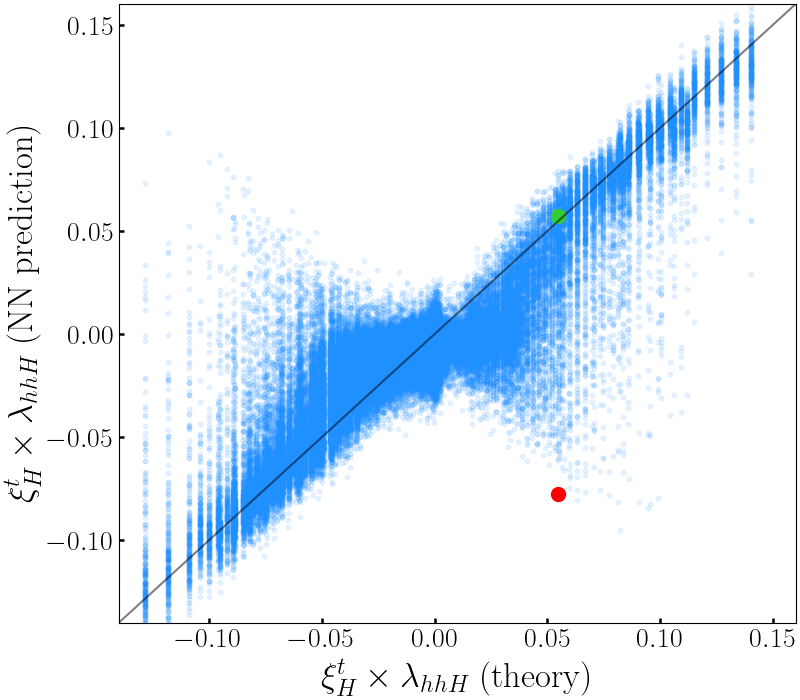}
\includegraphics[width=0.45\textwidth]{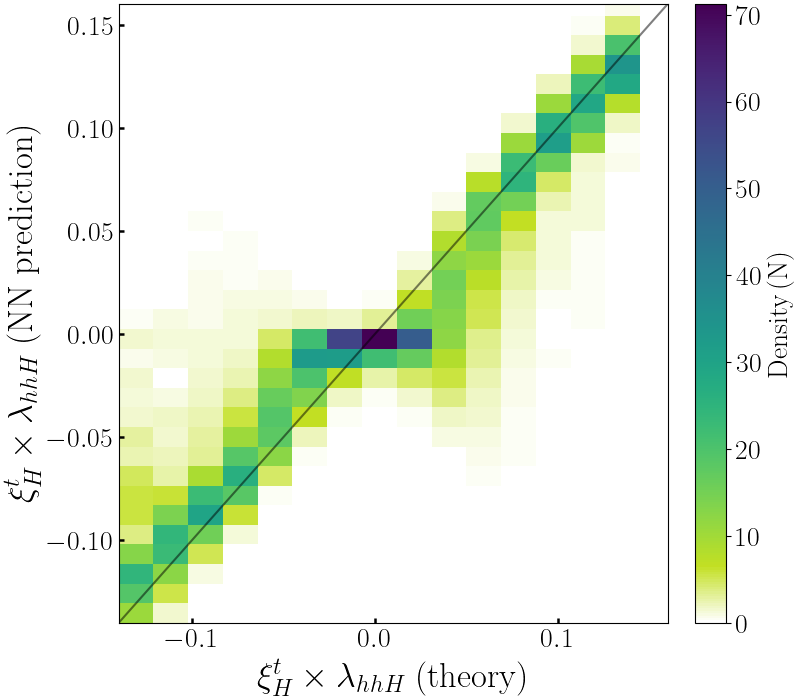}
\caption{Prediction of $\xila$ for points with $\MH = 450 \gev$ with the $\mhh$ distributions statistically smeared according to their respective
Poisson distributions. We show the results in a scatter plot (left) and a density plot (right), where in the latter case the color 
represents the density of points that falls into this grid. 
The points marked in green and red in the left plot are example points discussed in the text.}
\label{fig:results4b_1unc}
\end{figure}


Our final NN analysis was performed as follows. The NN was trained on the full set of \texttt{Model~I}. 
For each point in the plane we derived $2^{15}$ statistically smeared $\mhh$ distributions, for which we use the Poisson distribution
(see \refse{sec:stat_unc} for a definition of the smeared, and \refse{sec:stat} for the related smearing of the data). 
We then obtained the $\xilaNN$ values for 256 test $m_{hh}$ distributions that were also statistically smeared with their respective Poisson distribution. 
The result is shown in the form of a scatter plot as a function of the true $\xilaTH$
value in the left plot of \reffi{fig:results4b_1unc}. (The points marked in red and green in this plot are going to be discussed 
in more detail further below.) In comparison with the statistically
unsmeared case, see the left plot in \reffi{fig:2dvs3d}, a substantial broadening of the NN predicted $\xi_H^t \times \lambda_{hhH}$ values can be 
observed, resulting in an apparently strong degradation of the NN capability to find the correct value of $\xila$.
This, however, neglects the fact that most of the points that have a very different $\xilaNN$ value w.r.t.~the corresponding
$\xilaTH$ value are ``outliers'', and most of the predicted results obtained are still close to the diagonal. In order to 
demonstrate this, on
the right plot we present the density distribution of the points of the plot on the left, where we have 
normalized to 100 the density at each bin of $\xilaTH$. Thus, the density in one vertical strip 
corresponding to a given value of $\xilaTH$ adds up to 100. One can see that the highest 
density of points in the prediction falls into the diagonal line of  $\xilaNN = \xilaTH$. 
This means that despite there are some outliers, the NN prediction is reasonably~accurate for the majority of points.

\begin{figure}[ht!]
\centering
\includegraphics[width=0.45\textwidth]{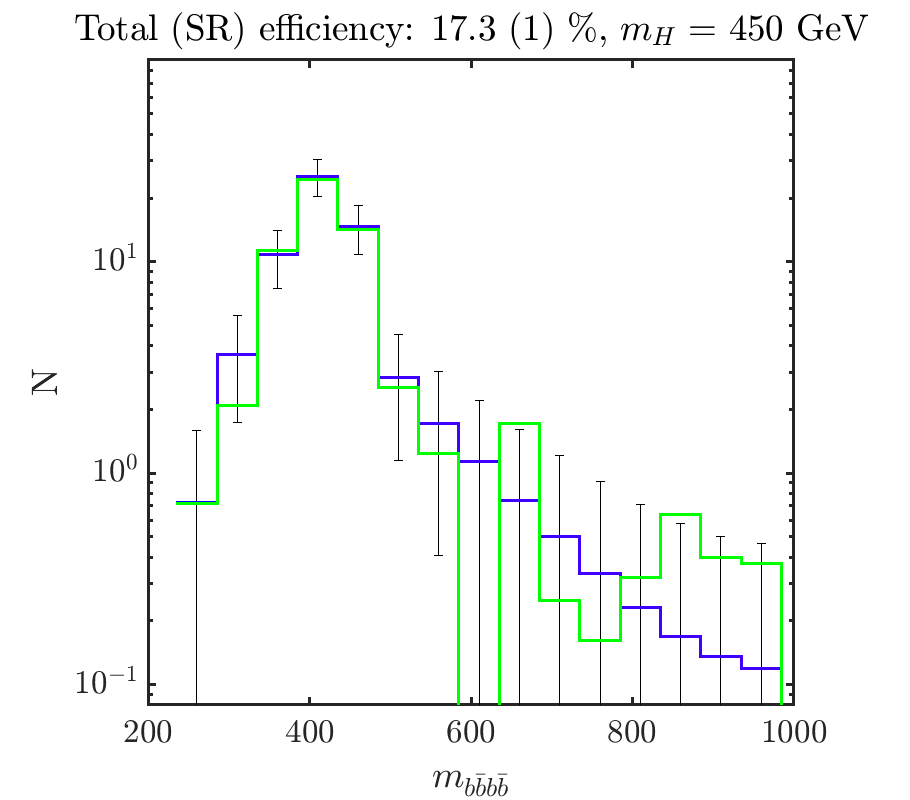}
\includegraphics[width=0.45\textwidth]{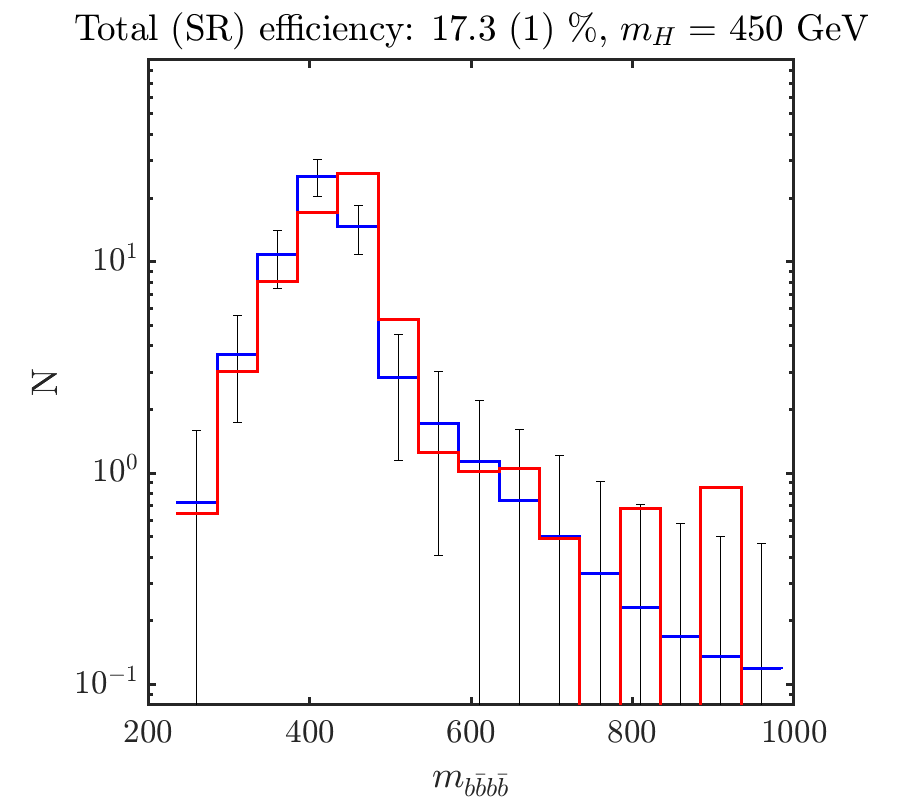}
\caption{Examples of invariant mass distributions of the point marked by a cyan cross in 
\protect\reffi{fig:prod_mH450} with the parameters given in \protect\refta{tab:bpoints}. 
The blue distributions are based on \xilaTH, and the vertical bars represent the statistical uncertainties.
The green (left) and red (right) distributions are randomly chosen  based on these uncertainties (see text).}
\label{fig:ex_4b_1unc}
\end{figure}

We have conducted a more detailed analysis of two distinct smeared distributions, both corresponding to the same benchmark point indicated by a cyan cross in \reffi{fig:prod_mH450}. Their predictions are represented in the left panel of~\reffi{fig:results4b_1unc}, with the green dot (``close''), which lies near the theoretical value, and the red dot (``far''), which deviates significantly from the theoretical prediction.
The parameters of the benchmark point are given in \refta{tab:bpoints}, together with the NN predictions
of the two smeared distributions.
We show these distributions in \reffi{fig:ex_4b_1unc}, 
where the blue $\mhh$ distribution is identical in both plots and has been calculated with $\xilaTH$.
As before, the vertical bars indicate the statistical uncertainty for each bin. 
Based on these uncertainties, the green (left plot) and red (right plot) $\mhh$ distributions are the ones that were randomly chosen for this point
to be fed to the trained NN to evaluate
$\xilaNN$. In the left plot we plot the distribution corresponding to the ``close'' point, which has a \xilaNN\ value close to \xilaTH, see \refta{tab:bpoints}, and which is represented by the green point in the left plot
of \reffi{fig:results4b_1unc}.
It can be seen that the key to get this good agreement in the NN prediction is the small
deviation of the randomly chosen distribution from the one based on \xilaTH, particularly in the five 
bins around the resonance  at $\mhh = 450 \gev$. This argument becomes clearer when looking at the right
plot, corresponding to the ``far'' point.
In this case the five bins closer to the resonance exhibit a shift in the relative height of the
true distribution w.r.t.\ the randomly chosen one, based on the statistical uncertainties.
These shifts produce event numbers with three out of five outside the respective $1\,\sigma$ interval
of the respective bin, resulting in the large difference of \xilaNN\ and \xilaTH. In this case,
the NN even predicts the wrong sign, indicated by the red point in the left plot of \reffi{fig:results4b_1unc}.

\begin{table}[ht!]
\begin{center}
\begin{tabular}{c|c|c|c|c}
      $\TB$ & $\CBA$ ($\SBA>0$) & \xila (th) &  \xila (NN) ``close'' & \xila (NN) ``far''\tabularnewline
     \hline
    4.38     &    -0.04     &  0.0547  &  0.0572 & -0.0775 \tabularnewline
\end{tabular}
\caption{Parameters for the point marked by the cyan cross in \protect\reffi{fig:prod_mH450}.
The value of $\xilaNN$ ``close'' (``far'') corresponds to the random distribution shown in green (red) in the left (right)
plot of \protect\reffi{fig:ex_4b_1unc}. (\xila (th) is short for \xilaTH, \xila (NN) is short for \xilaNN.)
The ``close'' point is marked in green in \reffi{fig:results4b_1unc}, and the ``far'' point is marked in red.
}
\label{tab:bpoints}
\end{center}
\end{table}

This raises the question whether the NN prediction of \xila\ can be improved by increasing the number of ``strongly'' statistically
Poisson smeared distributions. To analyze this, 
we trained the network on data with a larger uncertainty range (allowing $2\sigma$ error bars with the means of the Poisson distribution 
at the theoretical values of \xila) and validated it on data with a $1\sigma$ uncertainty. In this way we provide to the net more 
information about the points that deviate the most from the prediction, so that it can \textit{learn} more about them. 
The results can be seen in \reffi{fig:results4b_2sig}, where one can observe that the training extended to more strongly statistically
smeared distributions visibly improves the result. (A more quantitative analysis will be given below.) We also tested whether a further
increase of the statistical uncertainty in the training sample would further improve the NN prediction of \xila. However, we found that
the increase of the Poisson uncertainty by a factor of $\sim 2$ is close to the optimum value (see our more quantitative analysis below).

\begin{figure}[ht!]
\centering
\includegraphics[width=0.45\textwidth]{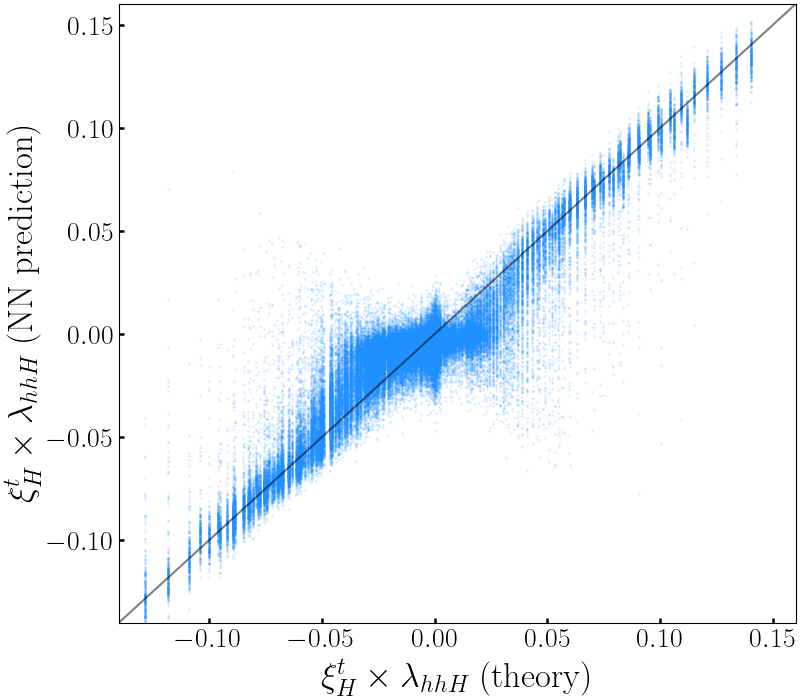}
\includegraphics[width=0.45\textwidth]{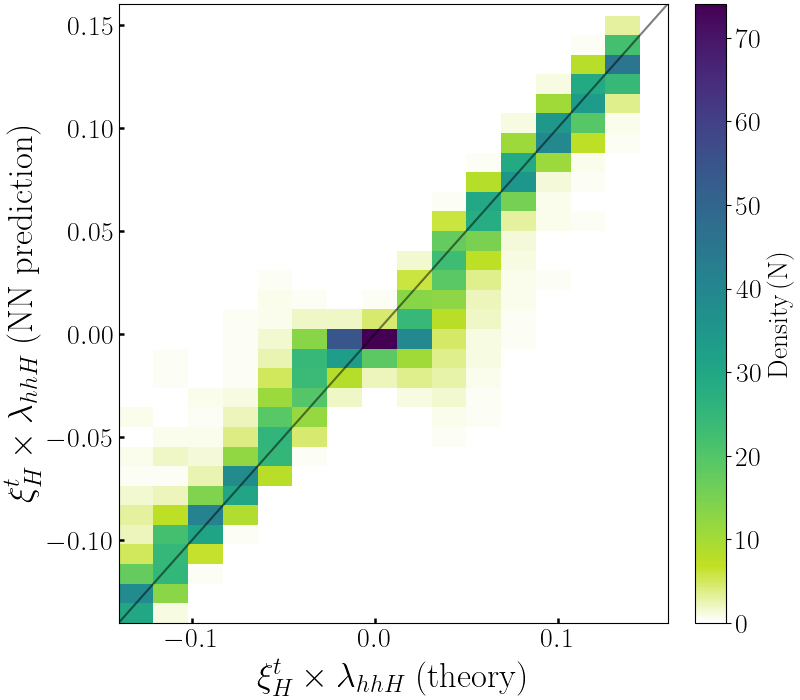}
\caption{Prediction of $\xila$ for points with $\MH = 450 \gev$ with the $\mhh$ distributions statistically smeared according to twice their respective
Poisson distributions. The color coding is as in \protect\reffi{fig:results4b_1unc}.}
\label{fig:results4b_2sig}
\end{figure}

\begin{figure}[ht!]
\centering
\includegraphics[width=0.45\textwidth]{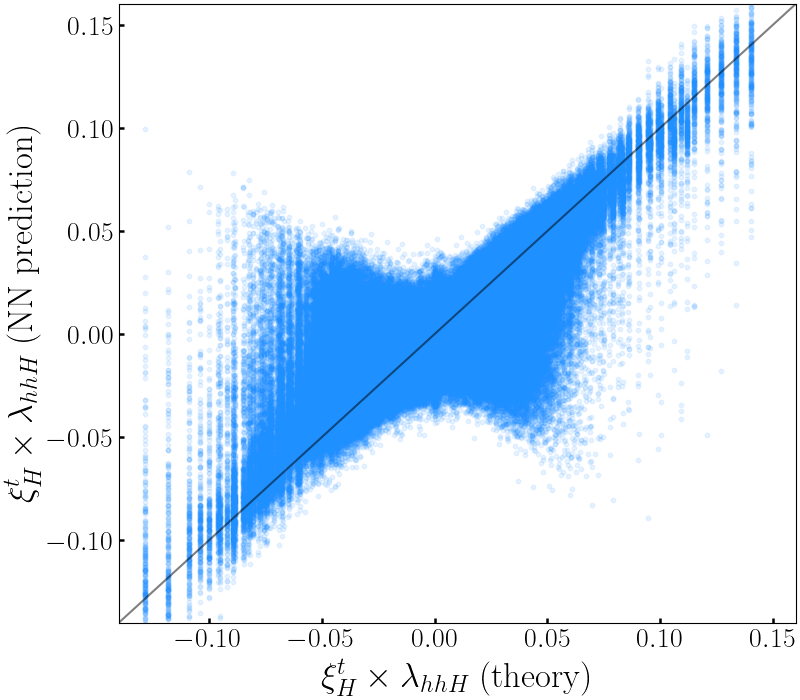}
\includegraphics[width=0.45\textwidth]{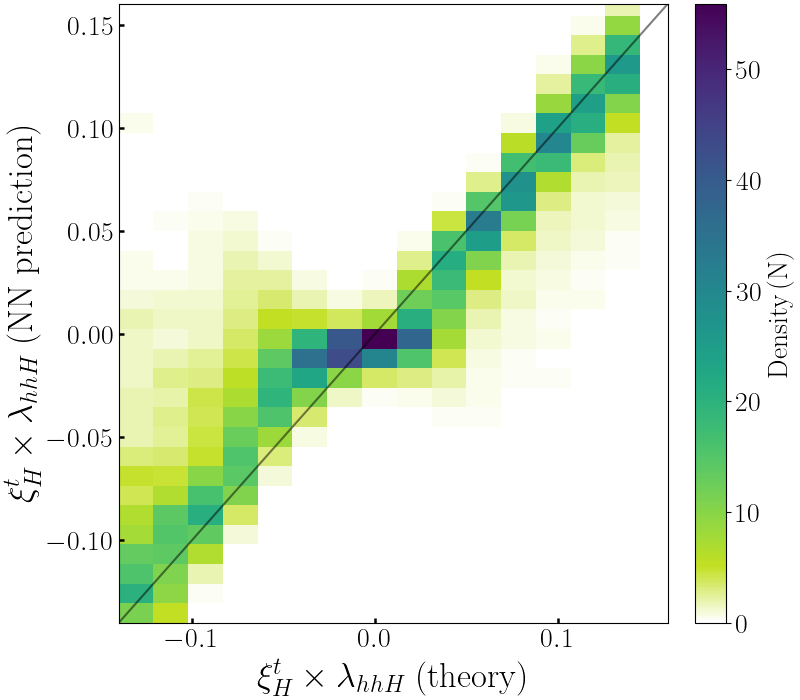}
\includegraphics[width=0.45\textwidth]{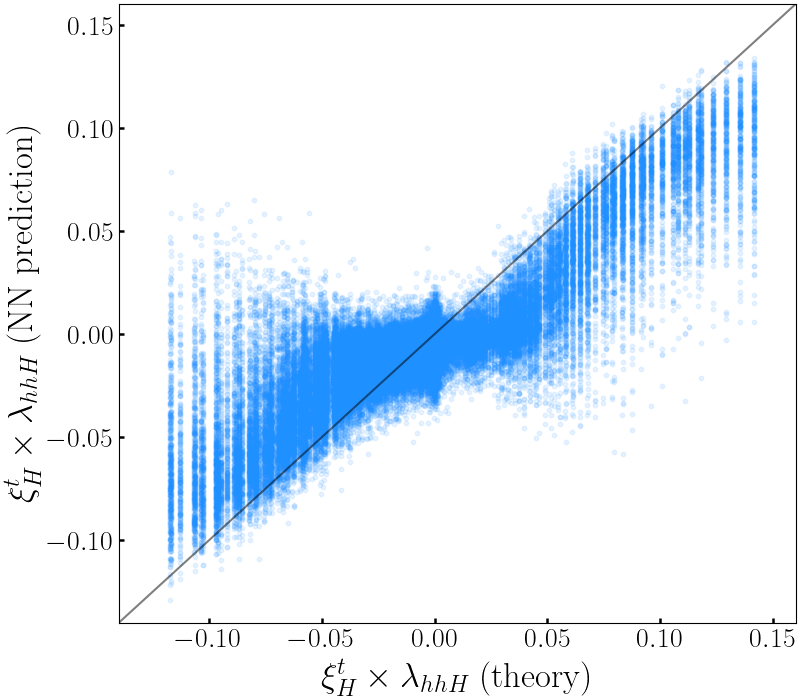}
\includegraphics[width=0.45\textwidth]{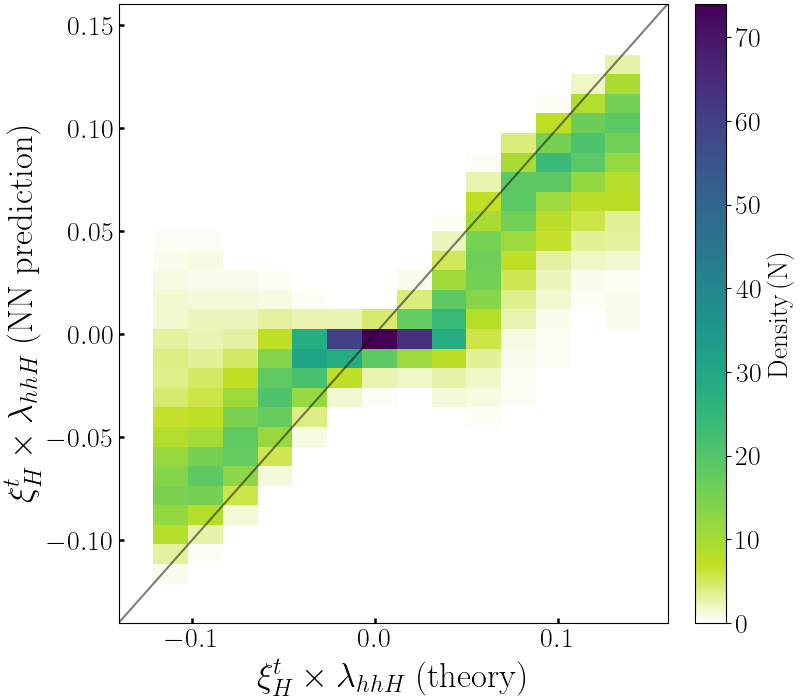}
\caption{Predictions for statistically smeared data. Upper row for free $\msq$ and lower row for $m_H = 443 \gev$ (as a test case) employing {\texttt Model~IV}.}
\label{fig:stat_mH_m12sq}
\end{figure}

As a further step, we also took into account the effect of statistical smearing on the dataset with different values of $m_{12}^2$ and $m_H$, i.e.\ \texttt{Model IV}, by changing one at a time. We show the results of each of these steps in \reffi{fig:stat_mH_m12sq}, where the upper row shows the effect of the uncertainty in the value of $\msq$ and the lower row shows the effect of the uncertainty in $m_H$ by including the statistically smeared samples in the NN training with the 3D dataset in the former case and the \texttt{Model IV} in the latter. Again the results are shown for 256 points in the scatter plot on the left side of the figure, and the corresponding density of points is colored on the right side.
Including the uncertainties related to the values of $\msq$ and $m_H$ clearly enlarges the span of the predicted values by the NN for a given 
value of $\xila$. However, the corresponding density plots still display a sufficiently good precision, which will be further discussed in the 
next subsection. 

\medskip
Our conclusion from this analysis is that the determination of the variable $\xila$ with the correctly trained NN
is possible, even taking into account all relevant experimental uncertainties (where we did not take into account the
-- to us unknown -- systematic experimental uncertainties). For this determination it is crucial to reduce the statistical
uncertainties, and the result depends strongly on the experimental efficiencies. But even with the low acceptance of
$\epsilon = 0.17 \times 0.01$ the right plot of \reffi{fig:results4b_1unc} demonstrates that an experimental determination of 
$\xila$ is possible.


\subsection{Comparison of NN and Statistical Analyses}

In this section, we compare the improvement
in the performance of the NN with respect to the classical MLE method on the original benchmark scenario. We furthermore give more 
quantitative results in the comparison of the results with different assumptions on the experimental uncertainties. To quantify this, we have used the 95\% confidence interval for the absolute error of \xila, AE95. We trained the NN and the MLE on an equal number of samples. The interpretation of AE95 is that the error in the prediction of \xila\ is at 95\% CL smaller than AE95, therefore, the smaller the value of AE95, the more reliable the prediction of the corresponding method. 

We find that the error is highly sensitive to the value of \xila\, as illustrated in \reffi{fig:compae95}. The left plot shows the AE95 for 
\xila\ computed using the MLE method, while the right plot displays the results from the NN. Both approaches reveal a similar overall trend: 
The largest errors occur near $\xila = \pm 0.05$ when using the NN ($\pm 0.07$ using the MLE method) and decrease for larger absolute values as well as for smaller ones. Additionally, the minimum error appears near zero, with a slight preference for negative values.

\begin{figure}[ht!]
\centering
\includegraphics[width=0.49\textwidth]{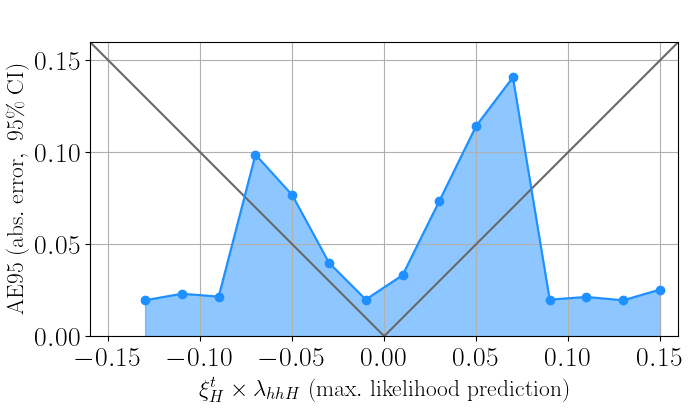}
\includegraphics[width=0.49\textwidth]{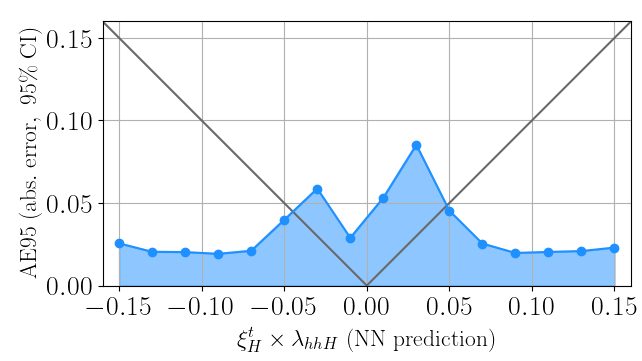}
\caption{ Comparison of the AE95 with MLE (left) and NN (right) depending on the value of $\xila$.}
\label{fig:compae95}
\end{figure}

We can define the ranges where the 95\% CL error is larger than the value itself as the crossing points between the blue and gray lines in \reffi{fig:compae95}, this comparison yields the \xila\ intervals: 

\begin{equation}\begin{split}
    \xila < -0.0755 \quad \text{or} \quad \xila > 0.0799 &\quad \text{for MLE},\\
    \xila < -0.0459 \quad \text{or} \quad \xila > 0.0450 &\quad \text{for NN (1$\sigma$)}.
\end{split}
\end{equation}
These intervals can be interpreted as the sensitivity ranges of \xila\ with each method.
It is clearly visible that the NN method yields a larger sensitivity to the BSM couplings.

\begin{figure}[ht!]
\centering
\includegraphics[width=0.55\textwidth]{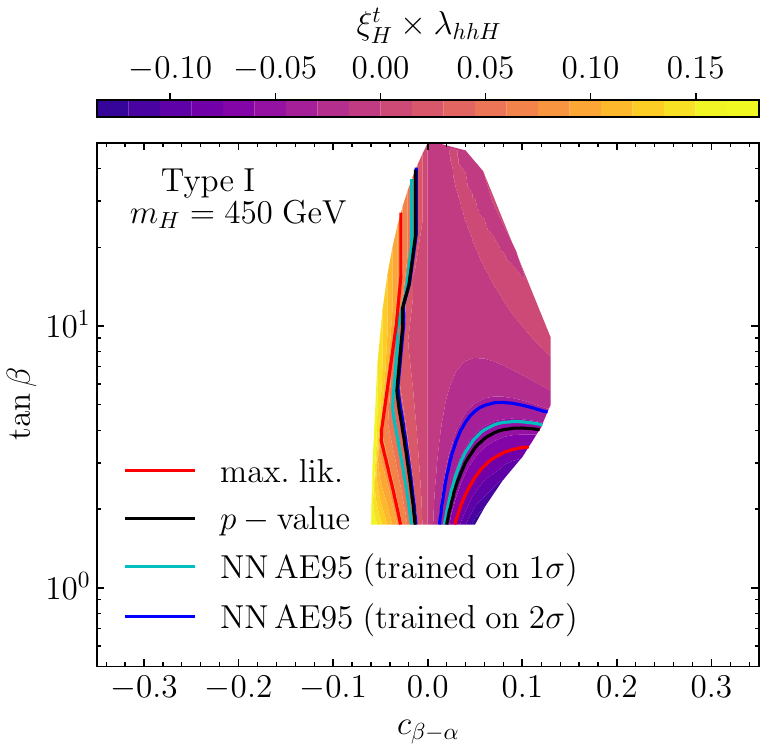}
\caption{Comparison between the NN and the classical MLE approach.}
\label{fig:comp_NN_stat}
\end{figure}

We can also use these intervals to compare the regions of the original plane that can be probed with both methods. This region is shown in 
\reffi{fig:comp_NN_stat}.
In this plot we display the original benchmark plane with the color coding indicating the magnitude of \xila. Inside the solid contour lines the 
determination of \xila\ would not be conclusive with each respective method.
We point out several observations: 1) The maximum likelihood (red contour) is worse than the NN (cyan contour) for the determination of \xila. 
2) The $p$-value (black contour) cannot perform a parameter estimation. It can only exclude the null hypothesis of the SM, and in this respect it is competitive 
with the NN. 
3) The NN can do both the hypothesis test and the parameter estimation simultaneously. While being comparable with the classical methods in the first 
case (hypothesis testing), it outperforms them in the second case (parameter estimation).
We have furthermore obtained the AE95 precision for the case that the Poisson uncertainty has been doubled for the training samples. 
In this case the AE95 limits for \xila\ are, 
\begin{equation}\begin{split}
\xila < -0.0334 \quad \text{or} \quad \xila > 0.0341 \quad \text{for NN (2$\sigma$)}.
\end{split}
\end{equation}
shown with a solid blue line in \reffi{fig:comp_NN_stat}, the blue line for positive \xila lies below the black line. The results are better than with other methods, in particular for negative values of \xila. 
Training the NN with points at $2\sigma$ from the mean value, provides more information about the points that deviate the most from the prediction, so that it can learn more about them. This explains how the NN becomes more robust. The optimal range (around $2\sigma$) means that there is a balance: too little noise limits learning from outliers, while too much noise makes it difficult for the NN to learn meaningful patterns. We have verified explicitly that training the net on values with even larger uncertainties worsens the results.

We therefore conclude that the NN is not only the best method for the BSM parameter determination, as it can do both hypothesis testing and parameter estimation in a more efficient way, it also has the potential to be improved with a dedicated analysis of the data.

\smallskip
Using the AE95 metric, we can also compare the performance of the net for the different training sets of data, taking into account various 
possible future experimental uncertainties. In this case the one corresponding to \texttt{Model I} and shown in \reffi{fig:results4b_1unc} 
is called dataset (1), the one including the $\msq$ as a free parameter shown in the upper row of \reffi{fig:stat_mH_m12sq} is named dataset (2),
and the one shown in the lower row of  \reffi{fig:stat_mH_m12sq}, the one corresponding to \texttt{Model IV}, we call dataset (3). We obtain the following results 

\begin{equation}\begin{split}
        \xila < -0.0459 \quad \text{or} \quad \xila > 0.0450 &\quad \text{for dataset (1)},\\
        \xila < -0.0386 \quad \text{or} \quad \xila > 0.0338 &\quad \text{for dataset (2)},\\
        \xila < -0.0503 \quad \text{or} \quad \xila > 0.0712 &\quad \text{for dataset (3)}.
\end{split}
\end{equation}

The respective AE95 for each value $\xila$ are shown in \reffi{fig:compNN_datas}. The key features are that interestingly, the net performs better for the dataset (2) than for dataset (1), i.e.~adding more free parameters of the model to the dataset improves its performance. It performs worse for the dataset (3) than for the dataset (1), specially for positive $\xila$ values, which suggests that an uncertainty in the measurement might be a source of error in the NN approach, which, however, can be brought under control by including a larger training sample with more $m_H$ values. We note in particular that the results for the dataset (2) appear to be better than the ones for dataset (3). The reason is that having $\msq$ as a free parameter enlarges the learning sample for the NN for values of $\xila$ close to the alignment limit, which are the ones for which the prediction is more challenging. In contrast, the values of $\xila$ further away from the alignment limit are easier to distinguish from the SM and these are present in the original scenario, as this is the one that allows largest mass splittings between the couplings while being consistent with the perturbative unitarity constraints. By changing $\msq$ we deviate from the optimal region for perturbative unitarity, and larger deviations in $\xila$ from the SM become disallowed. We conclude that a larger sample, even with more free parameters, would improve the determination of the $\xila$ by the NN.

\begin{figure}[h!]
\centering
\includegraphics[width=0.7\textwidth]{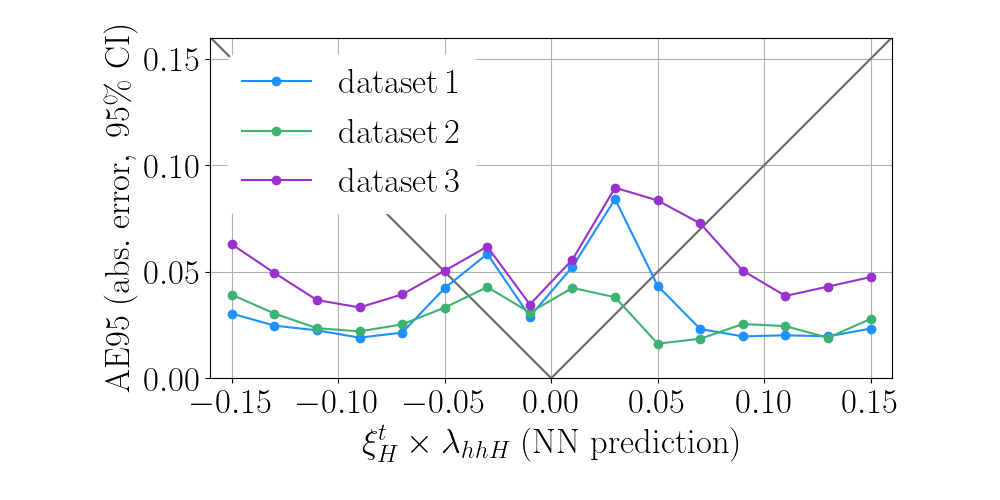}
\caption{
AE95 for the different datasets employed. Dataset 1 refers to the original benchmark plane, dataset 2 to the plane with $m_{12}^2$ free and dataset 3 is the one with the uncertainty in the value of $m_H$. }
\label{fig:compNN_datas}
\end{figure}



\subsection{Future Prospects}

The results in the previous subsection have  been obtained using the experimental efficiencies as given by
ATLAS~\cite{ATLAS:2022hwc}. However, it is conceivable that the efficiencies during the HL run of the LHC might 
improve due to a better knowledge of the systematic uncertainties with increasing luminosity. As a hypothetical
scenario we analyze here the level of improvement in the NN determination of \xila\ in the case that each of the two
efficiencies, $\epsilon_{\rm TOT}$ and $\epsilon_{\rm SR}$, are improved by a factor of two respectively. This increase of 
the number of events by a factor of four would yield a factor of~$1/2$ in the statistical uncertainty for each 
bin in the $\mhh$ distribution.

\begin{figure}[ht!]
\centering
\includegraphics[width=0.45\textwidth]{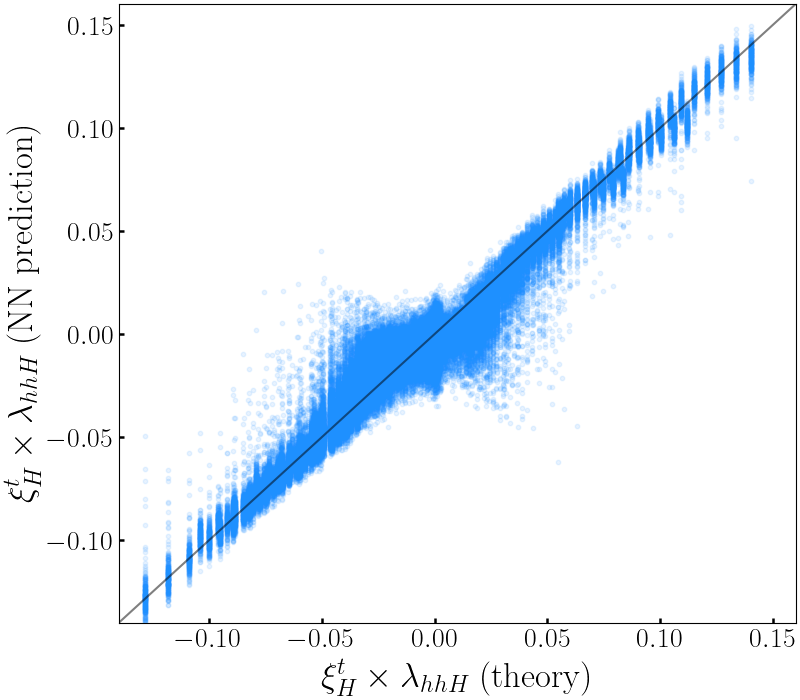}
\includegraphics[width=0.45\textwidth]{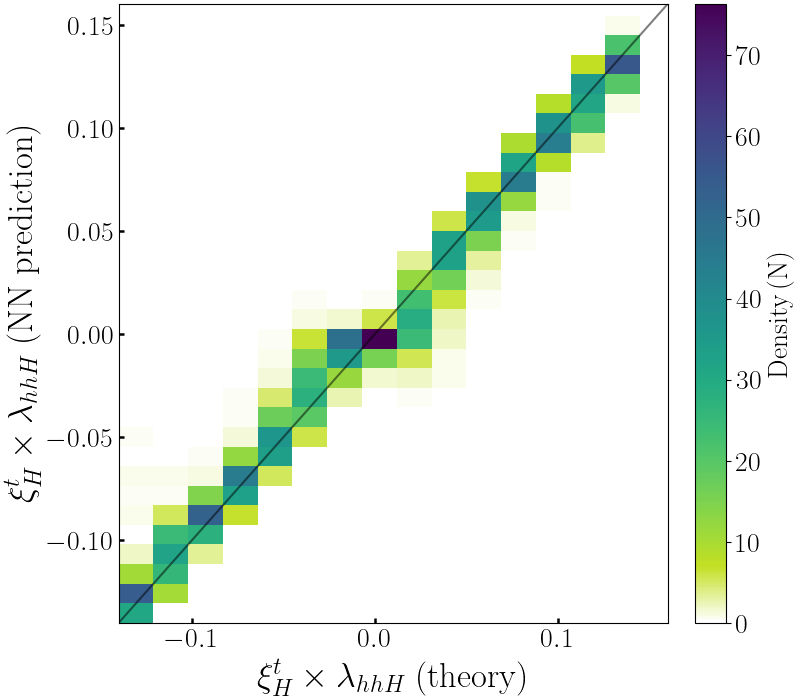}\\[2em]
\includegraphics[width=0.45\textwidth]{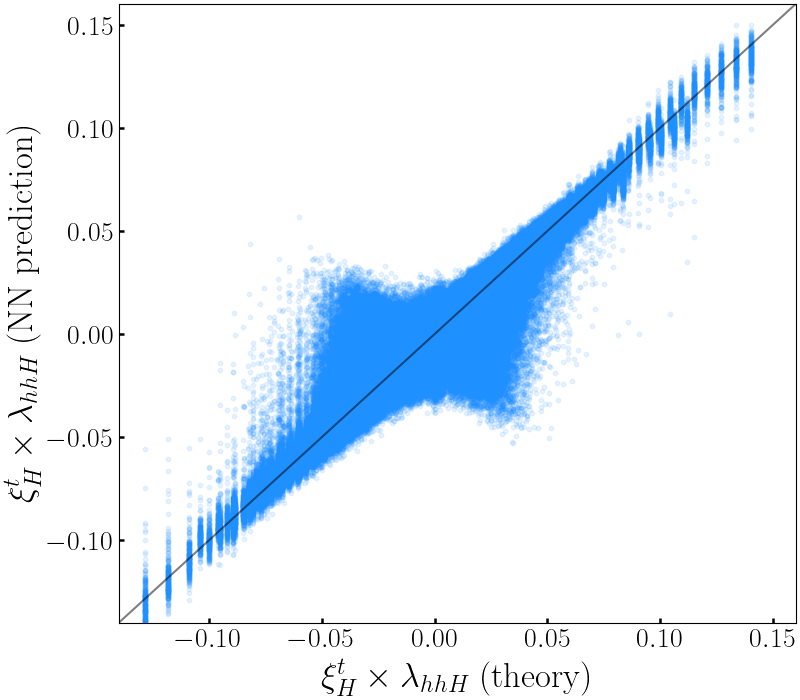}
\includegraphics[width=0.45\textwidth]{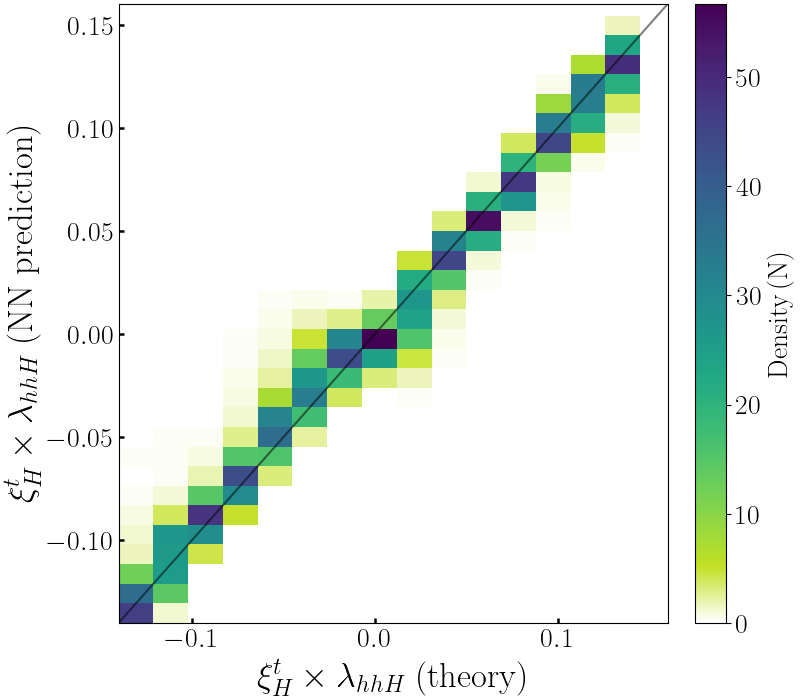}\\[2em]
\includegraphics[width=0.45\textwidth]{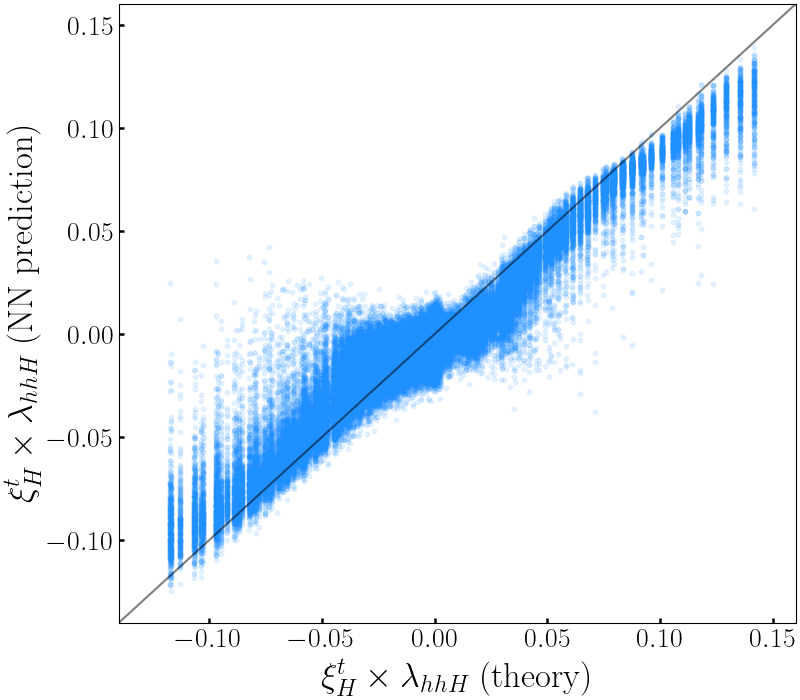}
\includegraphics[width=0.45\textwidth]{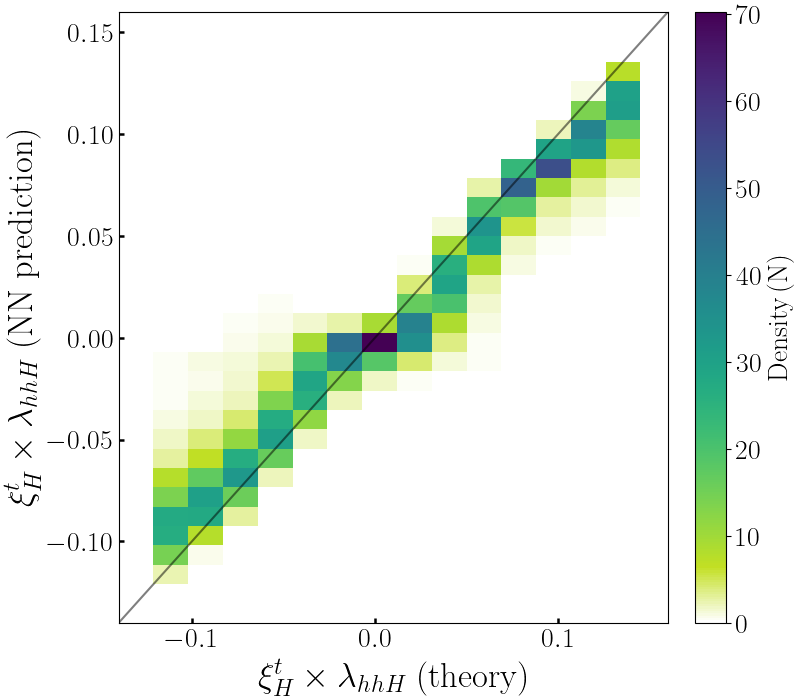}
\caption{Same as \reffi{fig:results4b_1unc} but with a factor 4 improvement in the total efficiency $\epsilon$ (see text).} 
\label{fig:results4b_1unc_4eff}
\end{figure}

In \reffi{fig:results4b_1unc_4eff} we show the results of our NN analysis assuming this improvement of the combined
efficiency by a factor of four, i.e.\ the same as in \refse{sec:resunc}, but with relative statistical uncertainties 
smaller by a factor of~$1/2$. We evaluate the improvements on the initial benchmark plane (first row), on the plane with the free $\msq$ (middle row) and, additionally, the free $m_H$ (lower row). We show the scatter plots on the left and the density plots on the right. All the scatter plots in the left side of \reffi{fig:results4b_1unc_4eff} show a substantial 
improvement compared to the corresponding (left) plots in \reffis{fig:results4b_1unc} and \ref{fig:stat_mH_m12sq}. 
The improvement becomes even more apparent
in the comparison of the density plots on the right of \reffi{fig:results4b_1unc_4eff} with the corresponding (right)
plots of \reffis{fig:results4b_1unc} and \ref{fig:stat_mH_m12sq}. 
With the efficiencies improved by a factor of two a NN determination of \xila\ at the level of 10-20\% appears to be feasible.


\section{Conclusions}
\label{sec:conclusions}

Di-Higgs production at the (HL-)LHC provides the unique opportunity to measure triple Higgs couplings to assess
the form of the Higgs potential. Beyond-SM Higgs sectors, as suggested by the open questions of the SM, not only introduce new Higgs bosons into the model,
but also the corresponding BSM THCs. Existing LHC studies of THCs concentrated on the THC of the Higgs boson at
$\sim 125 \gev$, $\lahhh$. A second ($\cp$-even) Higgs boson in the spectrum can contribute via 
an $s$-channel exchange to the process $gg \to hh$ and interfere with the $s$-channel exchange of the $h$ and the di-Higgs 
production via a top-quark box. This contribution is $\propto \xila$, the product of the top
Yukawa coupling of the new Higgs boson and its THC with two $125 \gev$ Higgs bosons.

The main goal of this
work is to assess how well a NN can determine \xila\ based on a ``realistic'' measurement of the $\mhh$
distributions from the process $gg \to hh$ at the HL-LHC. With ``realistic'' we refer to a 15\% smearing in $\mhh$
(corresponding to the anticipated experimental accuracy), as well as to a $50 \gev$ binning in $\mhh$ (corresponding
to the anticipated finite energy resolution).
As a concrete example, we focused on the 2HDM, providing naturally a second $\cp$-even Higgs boson, $H$.
We assumed a possible future scenario in which the $H$ will have been discovered with $\MH  \approx 450 \gev$, but the 
soft $\mathbb{Z}_2$ symmetry-breaking parameter $\msq$ is undetermined. The two mixing angles, parametrized by $\TB$ and $\CBA$, are
restricted by theoretical constraints, but also significantly by the LHC Higgs-boson rate measurements and the BSM Higgs-boson 
searches. These two mixing angles constitute the two main free parameters of our analysis.


For our analysis we employ a NN with the following characteristics: an input layer with 
16 nodes (corresponding the
input nodes for the binned $\mhh$ distribution of a certain \xila\ value) is followed by one layer with 
64 nodes. The final layer consists of one neuron that predicts the value of $\xila$. 
In general, the NN is trained on a subset of $\mhh$ distributions and then used to ``predict'' \xila\ corresponding to the
distributions it was not trained on.
In a first step we demonstrated that the result for \xila\ is not influenced by a variation in $\MH$ (assuming a possible
future HL-LHC accuracy for this mass), nor by a variation of $\msq$, the parameter of the 2HDM Higgs potential that
is most difficult to access.

Focusing on the $\CBA$--$\tb$ plane with $\MH$ and $\msq$ fixed, but neglecting statistical and systematical
uncertainties, we demonstrated that the NN can predict almost perfectly the value of $\xila$ from the (smeared and 
binned) $\mhh$ input. 

To properly take into account further uncertainties of the measurement, we incorporated the anticipated statistical uncertainties
in $\mhh$, assuming the $b\bar b\,b\bar b$ final state and $3 \iab$ of integrated luminosity (systematic uncertainties
are beyond the scope for this exploratory theoretical study). For a given theoretical prediction of $\mhh$ we evaluated
the statistical uncertainty based on the expected numbers of events and the anticipated experimental efficiencies in the
$b\bar b\,b\bar b$ final state. Taking these uncertainties as a Poisson $1\sigma$ uncertainty, we obtained for each 
point in the $\CBA$--$\tb$ plane $2^{15}$ Poisson smeared $\mhh$ distributions. For these ``realistic'' $\mhh$ distributions
the NN (trained only on the ``correct'' $\mhh$ distributions) then predicted the corresponding \xila\ value. We found 
that the highest density of \xila\ values in the prediction falls into the diagonal line of $\xilaNN = \xilaTH$. 
This demonstrates that despite there are some outliers, the majority of predictions are concentrated along the diagonal. In particular, the region around the identity line shows a point density exceeding 70\%, indicating that the neural network accurately captures the dominant trend between theory and prediction.
%
%

We have compared these results to the classical maximum likelihood estimation (MLE) methods for parameter estimation, and the results with NN proved to be better, with an even larger gain if the net is trained more on values that are outliers from the theoretical prediction. This not only shows the comparative improvement of machine learning versus classical statistics, it also highlights the potential for an improvement in a dedicated analysis.

Finally, we have analyzed the impact of a possible future improvement of the detector efficiencies in the HL run of the LHC. 
Our assumption of a factor 4 improvement in detector efficiencies would lead to a statistically significant determination of $\xila$
at the level of $10-20\%$.

If in the future data are available that confirm the existence of a BSM scalar particle in nature, a NN with a 
larger accuracy can be trained, by taking into account the future mass uncertainty range and by including higher-order loop
corrections to the process $gg \to hh$, and in particular to the involved THCs. We expect a more realistic and 
sophisticated analysis based on this future knowledge to outperform our results. Despite the fact that our analysis is model 
dependent, it is reasonable to assume that any model with a similar experimental signature that may then
be favored by experimental data will have good prospects for the determination of the couplings involved in the resonantly
produced Higgs boson pairs. In this way, NNs pave the way to determine the shape of the BSM Higgs potential.



\subsection*{Acknowledgments}
We thank Henning Bahl for his valuable input during the early stages of this work. We thank Jana Schaarschmidt for useful insights in the statistical uncertainties of the detectors and their efficiencies. 
K.R.\ thanks Panagiotis Stylianou for useful discussions and suggestions regarding neural networks.
The work of S.H.\ has received financial support from the
grant PID2019-110058GB-C21 funded by
MCIN/AEI/10.13039/501100011033 and by ``ERDF A way of making Europe'', 
and in part by by the grant IFT Centro de Excelencia Severo Ochoa CEX2020-001007-S
funded by MCIN/AEI/10.13039/501100011033. 
S.H.\ also acknowledges support from Grant PID2022-142545NB-C21 funded by
MCIN/AEI/10.13039/501100011033/ FEDER, UE.
The work of M.M. has been supported by the BMBF-Project 05H24VKB. K.R. acknowledges support by the
Deutsche Forschungsgemeinschaft (DFG, German Research Foundation) under Germany’s Excellence Strategy – EXC 2121 “Quantum Universe” – 390833306. This work has been partially funded
by the Deutsche Forschungsgemeinschaft (DFG, German Research Foundation) - 491245950.

\appendix
\section{Impact of Finite Grid Size of the Initial \boldmath{\mhh} Calculations}

We have used as the ``theoretical prediction'' the $\mhh$ distributions with 100~coordinates ranging from 
250~GeV to 1000~GeV, i.e.\ our initial bin size is 7.5~GeV. This value is precise enough for the overall 
\mhh\ distribution, but might be too coarse to capture well the peak around the heavy Higgs
resonance. A finer grid was not used initially because of computational efficiency: We are dealing with large 
parameter scans and need to run HPAIR in all the cases to obtain the theoretical prediction. 
In order to estimate the induced error of the ``large'' 7.5~GeV bin size on the \xila\ determination,
which we interpret as a further source of theoretical uncertainty, we have computed ten
benchmark points with an improved initial grid of 0.075~GeV. This is substantially finer than the one employed 
throughout this work by a factor of 100. For these points we obtained the predictions of $\xila$ with a NN trained on the points in the benchmark plane of \reffi{fig:prod_mH450}. The chosen points have $\TB = 1.74$ and are listed in 
\refta{tab:prespoints}.
One can observe that the NN prediction depends only very mildly on the initial grid size. We quantify the impact of the change of the grid size from 7.5~GeV to 0.075~GeV through
\begin{equation}
    \Delta \equiv \frac{|[\xilaNN_{0.075 \gev} - \xilaNN_{7.5 \gev}]|}{[\xilaTH]} \,.
\end{equation}
One can observe that $\Delta$ remains at the few percent level, 
i.e.\ the coarser grid size, which 
has been used to train the NN, yields nearly as good results as the finer one. From this we conclude that our analysis using the coarser grid size of 7.5~GeV captures all
the relevant effects. Any analysis in the future based on real experimental data should, of course, use the finest 
grid size possible by computational efficiency.

\begin{table}[ht!]
\begin{center}
\begin{tabular}{c|c|c|c|c}
      $\CBA$  & \xilaTH &  \xilaNN &  \xilaNN  & $\Delta$ [\%] \\
       &  & (7.5 GeV bin) &  (0.075 GeV bin) &  \tabularnewline
     \hline
     -0.06    &  0.168  &  0.166 &  0.166  &  $<$1\tabularnewline
     -0.05    &  0.140  &  0.139 &  0.139  &  $<$1\tabularnewline
     -0.04    &  0.111  &  0.111 &  0.111  &  $<$1\tabularnewline
     -0.03    &  0.083  &  0.083 &  0.080  &  4\tabularnewline
     -0.02    &  0.055  &  0.055 &  0.052  &  5\tabularnewline
     -0.01    &  0.027  &  0.027 &  0.024  &  12\tabularnewline
     0.00     &  0.000  &  0.000 &  0.000  &  $<$1\tabularnewline
     0.01    &  -0.027  &  -0.027 &  -0.025  &  8\tabularnewline
     0.02    &  -0.053  &  -0.053 &  -0.050  &  6\tabularnewline
     0.03    &  -0.079  &  -0.079 &  -0.077  &  3\tabularnewline
\end{tabular}
\caption{Predictions of the NN for $\mhh$ distributions with a finer grid. 
$\Delta$ quantifies the numerical impact on the \xilaNN\ determination from the grid size change 
(see text).}
\label{tab:prespoints}
\end{center}
\end{table}


\section{Details on the Neural Network Training}

\begin{figure}[!h]
\centering
\includegraphics[width=0.98\textwidth]{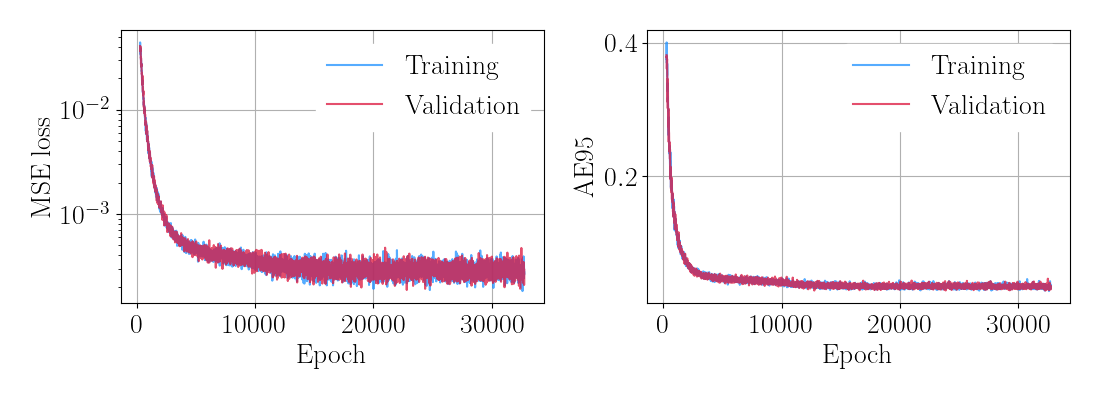}\\[2em]
\includegraphics[width=0.98\textwidth]{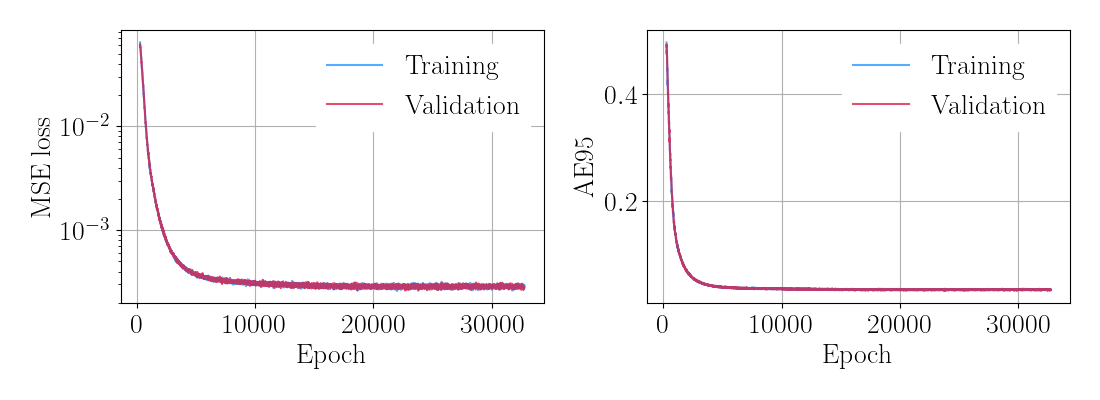}\\[2em]
\includegraphics[width=0.98\textwidth]{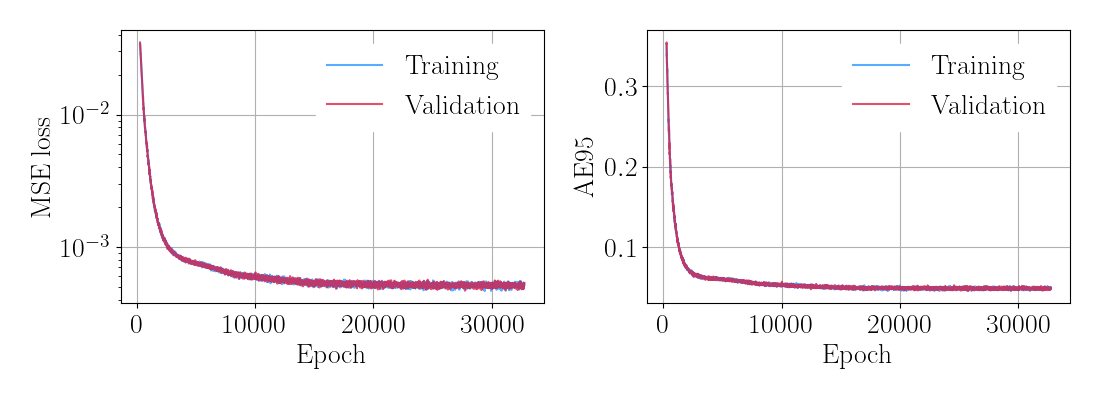}\\[2.5em]
\caption{NN prediction of $\xila$ for the different datasets: (1) upper, (2) middle and (3) lower row.}
\label{fig:results4b_1unc_loss}
\end{figure}

In \reffi{fig:results4b_1unc_loss} we show the MSE loss functions and the AE95 evolutions over the training epochs for the different models. The upper row corresponds to \texttt{Model I} (dataset (1)), the middle row to  \texttt{Model I} with $\msq$ free (dataset (2)), and the lower row to the \texttt{Model IV} (dataset (3)). These correspond to datasets with the original benchmark plane, with free $\msq$, and with an uncertainty in $m_H$, respectively.

Clearly, the training and validation losses closely track each other throughout the entire training process. There is no visible gap between them, and they decrease steadily over the epochs, which suggests that there is no overfitting.
We could consider an early stopping, as the MSE does not decrease significantly after $\sim 20$k epochs. However, we prefer to keep a longer learning rate to include further data, as each epoch learns from new Poisson smeared data.

We also mention here how we used the AE95 metric to decide on the network architecture.~As shown in \reffi{fig:compNN_archis}, there is no significant improvement in the ranges of \xilaNN\ when using a 2-hidden layer network with 128 neurons each or even a 4-hidden layer with 128 neurons each w.r.t~the simple 1-hidden layer network with 64 neurons applied in the present analysis.

\begin{figure}[h!]
\centering
\includegraphics[width=0.7\textwidth]{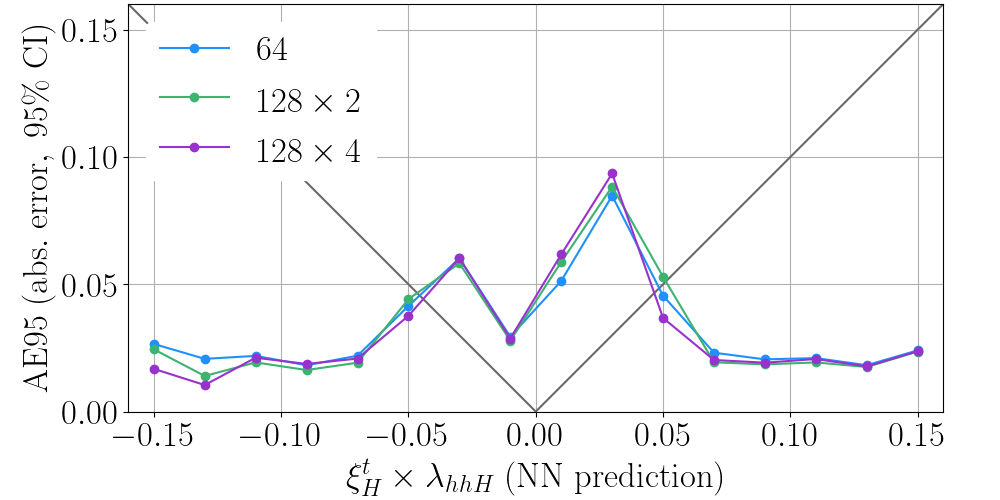}
\caption{ AE95 for different NN architectures. The one employed in this project is the simplest 1-hidden layer network with 64 neurons (blue), the more complex architectures include 2- and 4-hidden layer networks with 128 neurons each, shown in green and purple, respectively.}
\label{fig:compNN_archis}

\end{figure}

\bibliographystyle{JHEP}
\bibliography{lit}

\end{document}